\documentclass[a4paper,11pt]{article}
\pdfoutput=1

\usepackage[height=23.6truecm,width=16truecm,top=3.05truecm]{geometry}

\usepackage[skip=6pt,indent]{parskip}

\usepackage[utf8]{inputenc}
\usepackage{graphicx}
\usepackage{amsmath}
\usepackage{amssymb}
\usepackage{hyperref}
\hypersetup{colorlinks,citecolor=TokiwaIro,linkcolor=RuriIro,urlcolor=RuriIro,linktocpage}
\usepackage{cite}
\usepackage{braket}
\usepackage{bm}
\usepackage{bbm}
\usepackage{color}
\usepackage[svgnames,table]{xcolor}
\usepackage{float}
\usepackage{hhline}
\usepackage{multirow}
\usepackage{indentfirst}
\usepackage{latexsym}
\usepackage{mathtools}
\usepackage{cancel}
\usepackage{slashed}
\usepackage{colortbl}
\usepackage[mathscr]{euscript}
\usepackage{enumitem}
\usepackage{subcaption}
\usepackage{comment}
\usepackage{amsthm}

\usepackage{ulem}

\numberwithin{equation}{section}

\allowdisplaybreaks

\definecolor{RuriIro}{HTML}{1E50A2}
\definecolor{TokiwaIro}{HTML}{007B43}

\definecolor{dred}{rgb}{0.7,0.15,0.09}
\definecolor{dblue}{rgb}{0,0.12,0.64}
\definecolor{dgreen}{rgb}{0.2,0.51,0.19}
\definecolor{pegn}{rgb}{0.33,0.51,0.14}

\definecolor{kblue}{rgb}{0,0.48,0.73}
\definecolor{kred}{rgb}{0.73,0.25,0}
\definecolor{kgreen}{rgb}{0.48,0.73,0}

\definecolor{rgreen}{HTML}{7BAA17}
\definecolor{rred}{HTML}{AB1732}
\definecolor{rblue}{HTML}{007BBB}

\theoremstyle{plain}

\theoremstyle{remark}

\newcommand{\nn}{\nonumber}

\newcommand{\mc}{\mathcal}
\newcommand{\mr}{\mathrm}
\newcommand{\ms}{\mathsf}

\newcommand{\mbb}{\mathbb}

\newcommand{\msc}{\mathscr}
\newcommand{\mf}{\mathfrak}
\newcommand{\del}{\partial}
\newcommand{\ol}{\overline}
\newcommand{\dd}{\mathrm{d}}
\newcommand{\ee}{\mathrm{e}}
\newcommand{\iu}{\mathrm{i}}

\newcommand{\sld}{\slashed}

\newcommand{\Tr}{\mathop{\mathrm{Tr}}\nolimits}

\renewcommand{\Im}{\mathop{\mathrm{Im}}}

\newcommand{\diag}{\mathop{\mathrm{diag}}}

\newcommand{\SU}{\mr{SU}}
\newcommand{\U}{\mr{U}}

\newcommand{\YM}{\mathrm{YM}}
\newcommand{\sys}{\mathrm{sys}}
\newcommand{\env}{\mathrm{env}}
\newcommand{\fluid}{\mathrm{fluid}}
\newcommand{\loc}{\mathrm{loc}}
\newcommand{\ind}{\mathrm{ind}}

\newcommand{\gc}{\mathsf{g}}

\begin{document}

\begin{titlepage}

\begin{flushright}
\end{flushright}

\vspace{1cm}

\begin{center}

{\LARGE \bfseries
Bottom-up open EFT for non-Abelian gauge theory with dynamical color environment
}

\vspace{1cm}

\renewcommand{\thefootnote}{\fnsymbol{footnote}}
{%
\hypersetup{linkcolor=black}
Yoshihiko Abe$^{1,2,3}$\footnote[1]{yabe3@keio.jp}
and
Kanji Nishii$^{1,3}$\footnote[2]{kanji.nishii@keio.jp}
}%
\vspace{8mm}

{\itshape%
$^1${Graduate School of Science and Technology, Keio University, Yokohama, Kanagawa 223-8522, Japan}\\
$^2${Keio University Sustainable Quantum Artificial Intelligence Center (KSQAIC), Keio University, Tokyo 108-8345, Japan}\\
$^3${Quantum Computing Center, Keio University, 3-14-1 Hiyoshi, Kohoku-ku, Yokohama, Kanagawa, 223-8522, Japan}
}%

\vspace{8mm}

\end{center}

\abstract{
We develop a bottom-up open effective field theory (EFT) for non-Abelian gauge theories within the Schwinger--Keldysh formalism. 
Instead of integrating out the environment completely and starting from a nonlocal influence functional, we retain the slow environmental response variables explicitly and construct a local system-environment EFT. 
The environmental sector is described by a dynamical color-frame variable, St\"uckelberg-like field, and an associated color-current sector, which gives the nontrivial interactions and dissipation between the system and the environment. 
The resulting construction provides a gauge-covariant Markov embedding of nonlocal and non-Markovian color response. 
After integrating out the retained environmental variables with retarded boundary conditions, the reduced system theory acquires nonlocal dissipative kernels and stochastic sources. 
We show that the hard thermal loop response arises naturally as a particular realization of the retained environmental response. 
Our framework provides a local open-EFT description of color transport, memory effects, and fluctuation-dissipation structure in non-Abelian plasmas, and offers a systematic starting point for dissipative Yang--Mills EFTs with dynamical environments.
}

\end{titlepage}

\renewcommand{\thefootnote}{\arabic{footnote}}
\setcounter{footnote}{0}
\setcounter{page}{1}

\tableofcontents

\section{Introduction}

Real-time dynamics of nonequilibrium systems is one of the central subjects in the study of modern quantum field theory. 
Physical systems are rarely perfectly isolated. 
The low-energy and long-wavelength degrees of freedom of interest often interact with unobserved modes, thermal media, or high-energy hard modes.
As a result, their effective equations of motion contain dissipation, fluctuations, delayed response, and memory effects. 
These structures play essential roles in various fields such as hydrodynamics, quantum many-body systems, high-temperature plasmas, the early universe, heavy-ion collisions, and black-hole physics.

A natural framework for describing such nonequilibrium phenomena is the Schwinger--Keldysh (SK) formalism, which treats response functions, noise correlations, and initial-state dependence in a unified real-time contour formalism \cite{Schwinger:1960qe,Keldysh:1964ud}. 
The open-system viewpoint is encoded by the Feynman--Vernon influence functional \cite{Feynman:1963fq} and by standard dissipative models such as the Caldeira--Leggett model \cite{Feynman:1963fq,Caldeira:1982uj}. After environmental degrees of freedom are coarse-grained, the effective theory of the remaining system is generically non-unitary, dissipative, and stochastic. 
At the same time, microscopic unitarity survives as constraints on the effective theory, including SK reality, positivity, and dynamical Kubo--Martin--Schwinger (KMS) symmetry in modern non-equilibrium effective field theory (EFT) constructions \cite{Crossley:2015evo,Glorioso:2016gsa,Glorioso:2017fpd,Liu:2018kfw,Haehl:2015pja,Haehl:2015foa,Haehl:2018lcu}. 
In thermal equilibrium, these constraints are closely related to fluctuation-dissipation relations and linear response \cite{Callen:1951vq,Kubo:1957mj}, and can also be viewed as symmetry constraints on the SK action \cite{Sieberer:2015hba}. 
This symmetry-based open system EFT logic has been applied, for example, to spontaneous symmetry breaking, Nambu--Goldstone (NG) modes, and defect collective coordinates in dissipative systems \cite{Minami:2015uzo,Hongo:2018ant,Hongo:2019qhi,Landry:2019iel,Fujii:2021nwp}.

The environment is not necessarily an arbitrary bath added by hand. 
In many-body systems, thermal media, and plasmas, there is often a natural separation between the soft modes of interest and the microscopic degrees of freedom to be coarse-grained. 
If the latter are completely integrated out, the soft sector obtains a nonlocal and stochastic influence functional. 
On the other hand, long-lived modes associated with conserved or approximately conserved quantities are often more naturally retained explicitly in the low-energy EFT. 
Hydrodynamic variables are the standard examples \cite{Crossley:2015evo,Glorioso:2016gsa,Glorioso:2017fpd,Liu:2018kfw}. Transport coefficients, susceptibilities, relaxation rates, and noise kernels should then be understood as EFT data matched from the microscopic sector.

The present bottom-up EFT is not universal in precisely the same sense as hydrodynamics. We assume a separation between the low-frequency, long-wavelength dynamics retained in the EFT and the microscopic modes that have already been coarse grained out, so that the latter are encoded in local coefficients within the enlarged system-environment theory. Exact conservation laws determine the universal hydrodynamic variables, while approximately conserved quantities and other parametrically slow response channels provide additional candidates for explicit retention. The corresponding environmental variables are therefore selected according to the low-frequency response structure of the microscopic realization and are not, in general, uniquely determined by symmetry alone. Their field content may therefore depend on the microscopic realization, phase, and dynamical regime of interest. Once these variables are specified, gauge covariance, Ward identities, SK constraints, and dynamical KMS restrict their allowed couplings, while susceptibilities, transport coefficients, relaxation operators, and noise kernels remain matching data. The short-memory expansion introduced below is a further approximation, controlled by frequencies and momenta relative to the relaxation gap of the eliminated sector, and is not required for the enlarged local formulation itself.

This coarse-graining problem becomes more constrained in gauge theories. Because of gauge symmetry, conserved charges, and parallel transport, the effective response obtained after integrating out environmental degrees of freedom is generically nonlocal in spacetime and depends on the past history. 
The hard thermal loop (HTL) response in a high-temperature non-Abelian plasma is a representative example: after hard modes are integrated out, the soft Yang--Mills (YM) field acquires a nonlocal induced current, which admits both effective-action and kinetic descriptions \cite{Braaten:1989mz,Frenkel:1989br,Taylor:1990ia,Braaten:1991gm,Kelly:1994dh,Blaizot:1993zk,Blaizot:2001nr}. 
Its real-time fluctuation structure is naturally described in the SK formalism \cite{CaronHuot:2007nw}. 
Therefore, in an open EFT of non-Abelian gauge fields, it is not sufficient to simply add dissipation and noise in a gauge-covariant form. One needs a construction compatible with fundamental requirements such as Ward identities, gauge covariance, positivity, and KMS relations.

The starting point of this paper is to keep the low-energy environmental response explicitly. 
This gives a local system-environment EFT. 
The same logic appears naturally in dissipative open systems with dynamical gravity. 
Since gravity couples universally to all degrees of freedom, integrating out the dissipative environment completely obscures the compatibility of the Einstein equation with total energy-momentum conservation and diffeomorphism invariance. 
In gravitational open EFT, it is therefore useful to model the environment as a fluid sector and to construct a local system-environment EFT including its stress tensor \cite{Lau:2024mqm}. 
Related open EFT constructions with gauge or spacetime symmetries have also appeared in other contexts \cite{Salcedo:2024nex,Salcedo:2024smn,Salcedo:2025ezu,Christodoulidis:2025ymc,Colas:2025ind,Christodoulidis:2025vxz,Salcedo:2026cqb,Yoshimura:2026vil,Bu:2026mxr}.

\begin{figure}[t]
	\centering
	\includegraphics[width=0.48\textwidth]{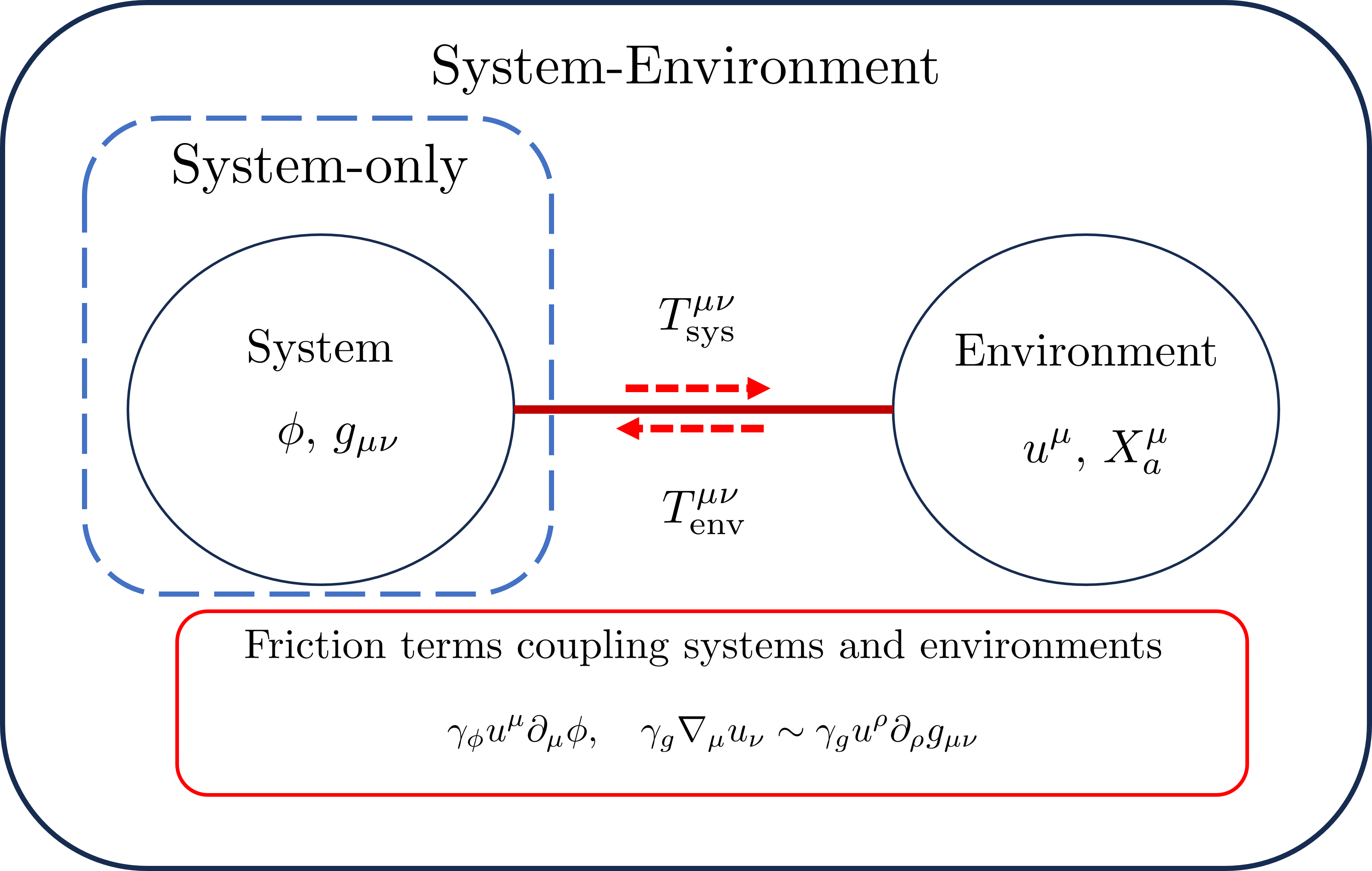}
	\quad 
	\includegraphics[width=0.48\textwidth]{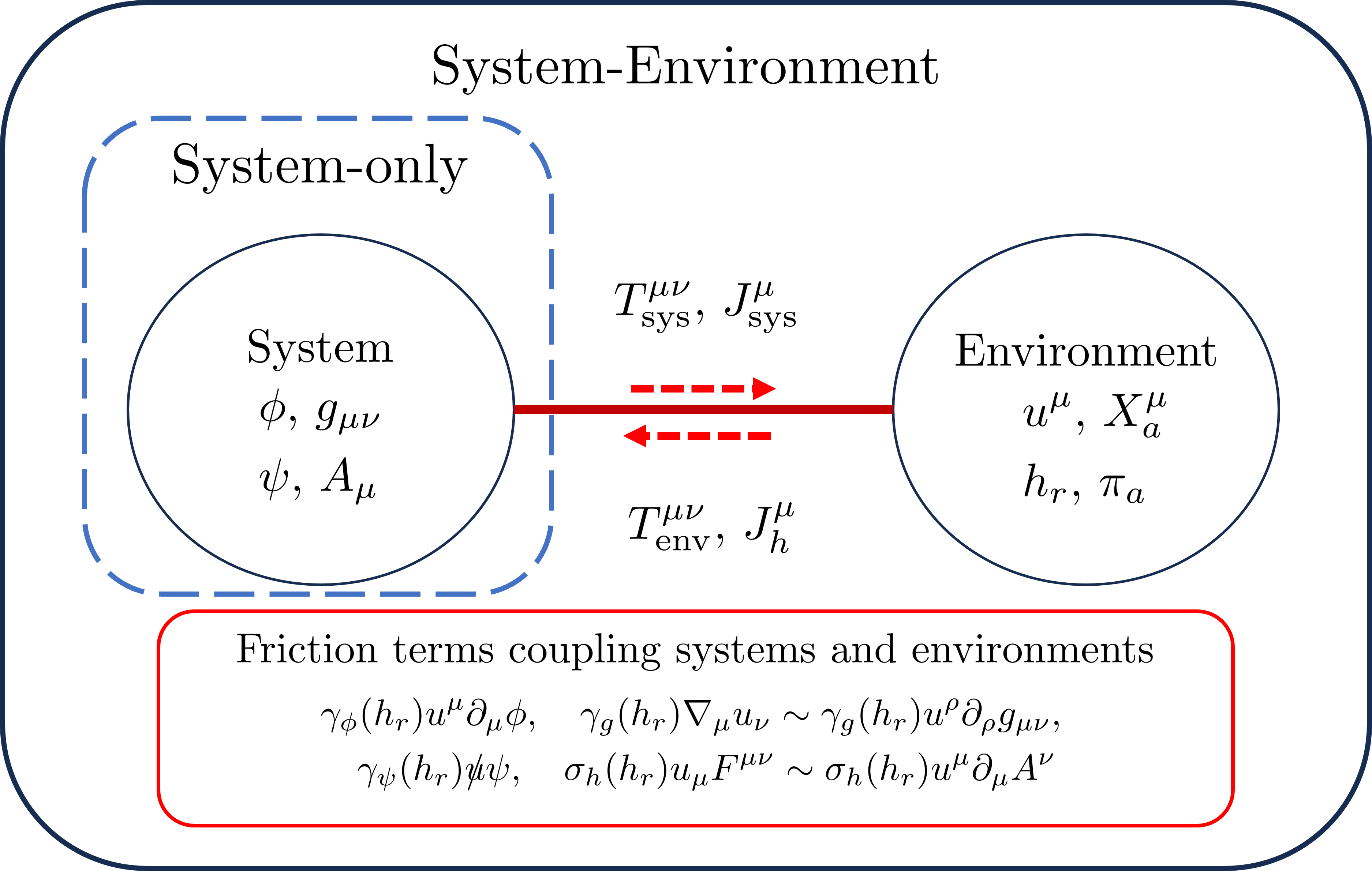}
    \caption{
    Schematic pictures of the system-environment construction for the gravitational (left) and the non-Abelian (right) open EFT.
    The system fields exchange the energy-momentum and color charge with retained environmental sector described by the dynamical environmental variables.
    The environmental currents are required to close the Ward identity and provide the nonvanishing charge exchange between the system and the environment.
    After integrating out the retained response variables with retarded boundary conditions, the reduced system-only description acquires nonlocal dissipative and stochastic responses.
    However, the system-environment construction gives the local EFT description, which is regard the Markovian embedding.
    }
    \label{fig:summary}
\end{figure}

In this paper we extend the local system-environment description of Ref.~\cite{Lau:2024mqm} to non-Abelian gauge theory. 
Instead of taking as the starting point a nonlocal influence functional obtained after completely integrating out the environment, we retain the environmental response arising nonlocal and non-Markovian effects.
In this sense, our construction may be viewed as a gauge-theoretic Markov embedding of an effective nonlocal memory kernel into an enlarged local dynamics \cite{Nakajima:1958pnl,Zwanzig:1961zz,Mori:1965oqj}. 
If one looks only at the system, the response contains memory kernels and is non-Markovian. 
In the enlarged theory including the retained environmental variables, however, the dynamics becomes local and Markovian. 
Nonlocality and memory appear only after the environmental response is solved using retarded Green functions and encoded back to the system-only description.
Fig.~\ref{fig:summary} shows the schematic picture of our EFT construction and the correspondence to the dissipative gravitational EFT.

For non-Abelian gauge theory, the environmental response required for this Markov embedding is an environmental color current that closes the color Ward identity. 
Just as the environmental stress tensor carries the energy and momentum dissipated from the system in the gravitational case, the environmental color current should carry the color charge exchanged with the YM field. 
This current is not an external bath inserted by hand, but an effective long-wavelength response of the coarse-grained hard sector or of colored degrees of freedom in the medium. 
To describe this response locally and gauge-covariantly, we introduce an environmental color frame and its corresponding $a$-type variable, which implements the Ward identity for the environmental color current. 
This structure parallels the St\"uckelberg-like variables that compensate relative diffeomorphisms in gravitational open EFT. 
The color frame alone, however, is not yet a dissipative color environment: the environmental sector contain a color current with color-charge storage, susceptibility, conductivity, relaxation, and noise. 
This makes it natural to regard the environment as a color-carrying transport sector, rather than as a purely compensating set of fields.  
Earlier work on colored particles and non-Abelian fluid dynamics provides a natural language for this color-fluid aspect \cite{Wong:1970fu,Bistrovic:2002jx,Fernandez-Melgarejo:2016xiv}, while effective descriptions of ultrasoft color conductivity and stochastic YM dynamics capture closely related long-wavelength transport physics in hot non-Abelian plasmas \cite{Bodeker:1998hm,Arnold:1998cy,Bodeker:1999ey,Blaizot:1999xk,Bodeker:2000da,Bodeker:2002gy,Arnold:2002zm}.

After the retained environmental response is solved using retarded Green functions and eliminated, the reduced description of the soft gauge field contains a nonlocal induced color current and a color-valued stochastic source.  
The Wilson-line structure of the reduced kernel is therefore not postulated as a bilocal term in the local action, which is generated by gauge-covariant retarded propagation in the environmental sector. 
The HTL response provides a useful benchmark for this mechanism.  
In its standard kinetic description, the long-wavelength color response of hard modes can be represented by a local covariant transport equation.  
In our language, this response is a retained environmental variable; eliminating it using retarded Green functions gives the usual HTL-type nonlocal induced current.

The organization and main results of this paper are as follows.  
In Sec.~\ref{sec:SK-formalism}, we review the general structure of open system EFT in the SK formalism and the St\"uckelberg-like construction in dynamical gravity.  
In Sec.~\ref{sec:NA-gauge}, we construct the environmental color frame, the corresponding $r$-type/$a$-type sources, and an environmental color-current sector that closes the covariant Ward identity.  
This gives a local system-environment description of color-charge storage, color transport, relaxation, noise, and gauge-covariant damping of colored matter.
In Sec.~\ref{sec:HTL-matching}, we introduce a velocity-resolved hard response and show that its retarded elimination gives an HTL-type nonlocal response.  
Sec.~\ref{sec:conclusion} is devoted to our summary and discussion of future directions.

\paragraph{Related work.}
Kaplanek, Mylova, and Tolley have recently developed a top-down Schwinger--Keldysh formulation of gauge theories in covariant gauges, with explicit BRST control of gauge fixing, influence functionals, general initial states, and finite-time boundary terms \cite{Kaplanek:2025moq,Kaplanek:2026kpp}.  
They also discuss HTL effective theory and open EFTs in a Higgs phase.

Some of their Higgs-phase formulas are structurally close to the expressions used below.  
In the Keldysh limit, for instance, one encounters the covariant combination
\[
	A_{a\mu}+D_\mu[A_r]\pi_a
\]
together with kernels depending on retarded Stückelberg data.  
The physical interpretation, however, is different.  
In their construction, the Stückelberg fields provide a gauge-covariant rewriting of a Higgs-phase open EFT and make the retarded gauge symmetry and Keldysh BRST symmetry manifest. 
In the present paper, by contrast, the gauge symmetry is unbroken.
The color-frame and $a$-type variables introduced below are not Higgs or Goldstone Stückelberg fields, but retained environmental response variables whose role is to complete the unbroken color Ward identity and to encode color relaxation, colored noise, and hard-loop memory.

Thus the two approaches address different layers of the same general problem.
Their work gives a first-principles SK/BRST control of gauge-theory influence functionals, while ours constructs a Ward-complete semiclassical Markov embedding of the environmental color sector.

\paragraph{Conventions.}

In this paper, we work in the mostly-plus convention for the (local) Lorentizan metric and the following Clifford algebra
\begin{align}
	\eta_{ab} = \diag(-1,+1,\ldots,+1) \,,
	\qquad 
	\{\gamma^a, \gamma^b\} = 2 \eta^{ab}\,,
	\label{eq:intro-metric-conventions}
\end{align}
where $a,b,\ldots,$ are indices of the local Lorentz frame.
It is noted that $\gamma^0$ is anti-hermitian in this convention $(\gamma^0)^2 =- \bm{1}$.
We define the Dirac conjugate of the Dirac spinor $\psi$ by 
\begin{align}
	\bar{\psi} \coloneqq \iu \psi^\dagger \gamma^0\,.
	\label{eq:intro-dirac-conjugate}
\end{align}
The curved spacetime metric $g_{\mu\nu}$ is related to $\eta_{ab}$ by the vielbein or the frame field $e^a{}_\mu$ as 
\begin{align}
	g_{\mu\nu} = \eta_{ab} e^a{}_\mu e^b{}_\nu\,,
	\label{eq:intro-metric-vielbein}
\end{align}
and the volume factor is written as 
\begin{align}
	\sqrt{-g} = \sqrt{- \det g_{\mu\nu}} = \det e^a{}_\mu = e\,.
\end{align}
The gamma matrix in the curved spacetime is introduced by $\gamma^\mu \coloneqq \gamma^a e_a{}^\mu$.
We use the slash notation for the contraction with the gamma matrix, e.g. $\sld{A} = \gamma^\mu A_\mu$.

In the open system, let us introduce a future-directed unit time-like environmental velocity $u_\mu$ $(u_\mu u^\mu = -1)$ to specify the time-direction of the dissipation.
The projector onto spatial surface orthogonal to $u^\mu$ is defined by 
\begin{align}
	\Delta^{\mu\nu}_u \coloneqq g^{\mu\nu} + u^\mu u^\nu\,,
	\qquad 
	\Delta^{\mu\nu}_u u_\nu = 0\,.
	\label{eq:intro-u-projector}
\end{align}
This $u^\mu$ is a slow environmental datum.
The dissipative terms such like $u^\mu \nabla_\mu \phi$ for a scalar field become Lorentz covariant once the frame specified by $u_\mu$ is fixed, but they are not invariant under the vacuum Lorentz group action.
In this paper, we work at the leading order of $u^\mu$ in the effective actions and treat this velocity as constant over the response time in the system.
The electric field measured by the above frame characterized by $u^\mu$ is introduced 
\begin{align}
	E^\mu \coloneqq F^{\mu\nu} u_\nu\,,
	\qquad 
	u_\mu E^\mu = 0\, ,
\end{align}
where $F_{\mu\nu}$ is the field strength of the gauge field $A_\mu$.
The details of our convention for non-Abelian gauge field are fixed in Sec.~\ref{sec:relative-gauge-frame}.
In the local rest frame of the flat spacetime $u^\mu = (1, \bm{0})$, we find that 
\begin{align}
	\Delta^{00}_u = 0\,, 
	\qquad 
	E^0 = 0\,,
	\qquad 
	\sld{u} = - \gamma^0\,.
\end{align}

\section{Schwinger--Keldysh formalism and bottom-up open EFT}
\label{sec:SK-formalism}

The EFT construction discussed in this paper is based on a bottom-up open-EFT viewpoint. Instead of considering only a dissipative system theory as fundamental one by including quantum effects from the environment, we study a local system-environment EFT by keeping slow environmental variables such as hydrodynamic modes, conserved density modes, or symmetry related NG type variables~\cite{Minami:2015uzo,Crossley:2015evo,Glorioso:2017fpd,Glorioso:2016gsa,Liu:2018kfw,Hongo:2018ant,Hongo:2019qhi}. Here, by \emph{retained} we mean that these slow environmental variables are kept explicitly in the local system-environment EFT. This approach provides an Markovian embedding of the (non-Markovian) system dynamics after integrating out the environment. This fact is also said that the retained environmental variables describes the dissipation between the system and the environment while keeping the symmetry structure manifest, and the integration of them produces the nonlocality in the system theory. The nonlocality or non-Markovian property appear as the memory kernel and the nonlocal dissipative response. A realization in a gravitational system based on this viewpoint was given in Ref.~\cite{Lau:2024mqm}.

This section fixes the SK language and system-environmental description of open-EFT, which are main tools in the following analysis.
Sec.~\ref{subsec:SK-constraints} reviews the doubling of fields, SK unitarity, the $r$-$a$ expansion, and the relation between response and noise.  
In Sec.~\ref{subsec:SK-grav}, we recall the open-EFT construction of the dissipative gravitational system in Ref.~\cite{Lau:2024mqm} as the guiding example for the non-Abelian theory.

\subsection{System sector, environmental sector, and SK constraints}
\label{subsec:SK-constraints}

We first recall the general SK-EFT description of nonequilibrium real-time dynamics.
Let fields $\Phi$ be system variables and $Y$ be those in environment.
The SK formalism doubles both sectors into forward and backward copies denoted by $+$ and $-$ so that the real-time response and stochastic fluctuations are encoded in a single functional with sources $J_\pm$ coupled to $\Phi_\pm$
\begin{align}
	Z[J_+, J_-] = \int \msc{D}\Phi_\pm \msc{D} Y_\pm \,\rho_0[\Phi_+, Y_+; \Phi_-, Y_-] \exp \left(
		\iu S[\Phi_+, Y_+] - \iu S[\Phi_-,Y_-] 
        + \iu \int \dd^d x \left( J_+ \Phi_+ - J_- \Phi_- \right)
	\right)\,,
	\label{eq:sec2-full-SK}
\end{align}
where $\rho_0[\Phi_+, Y_+; \Phi_-, Y_-]$ denotes the density matrix of the initial state.
$S[\Phi, Y]$ is the action for the system and the environment, and we assume that it is decomposed to 
\begin{align}
	S[\Phi, Y] = S_\Phi[\Phi] + S_Y[Y] + S_{\mr{int}}[\Phi, Y]\,,
\end{align}
where $S_\Phi[\Phi]$ and $S_Y[Y]$ are the actions for the system and environmental variables, respectively.
$S_{\mr{int}}[\Phi, Y]$ is the interaction between them.
We obtain the influence functional \cite{Feynman:1963fq} by integrating out the environmental variable $Y_\pm$ as 
\begin{align}
	\exp \iu I_{\mr{IF}}[\Phi_+, \Phi_-] \coloneqq \int \msc{D}Y_{\pm}\, \rho_Y[Y_+, Y_-] \exp \left(
		\iu S_Y[Y_+] - \iu S_Y[Y_-] + \iu S_{\mr{int}}[\Phi_+, Y_+] - \iu S_{\mr{int}}[\Phi_-, Y_-]
	\right)\, .
	\label{eq:sec2-influence-functional}
\end{align}
$\rho_{Y,0}$ denotes the initial density matrix of the environment.
Here, we assume a factorized initial density operator as $\hat{\rho}_0 = \hat{\rho}_{\Phi,0} \otimes \hat{\rho}_{Y,0}$.\footnote{If the initial system-environment correlations exist, they would give additional initial time contributions to the influence functional.}
The functional \eqref{eq:sec2-full-SK} becomes by using this influence functional as 
\begin{align}
	Z[J_+, J_-] = \int\msc{D}\Phi_\pm \,\rho_{\Phi,0}[\Phi_+,\Phi_-] \exp\left(
		\iu S_\Phi[\Phi_+] - \iu S_\Phi[\Phi_-] + \iu I_{\mr{IF}}[\Phi_+,\Phi_-] 
        + \iu \int \dd^dx \left(J_+ \Phi_+ - J_- \Phi_- \right)
	\right)\,,
\end{align}
and we introduce the system effective action by 
\begin{align}
	I_{\mr{eff}}[\Phi_+, \Phi_-] \coloneqq S_\Phi[\Phi_+] - S_\Phi[\Phi_-] + I_{\mr{IF}}[\Phi_+,\Phi_-]\,.
\end{align}
The microscopic unitarity of the underlying closed system implies the standard SK constraints on this effective action 
\begin{align}
	I_{\mr{eff}}[\Phi, \Phi] = 0\,,
	\qquad 
	I_{\mr{eff}}[\Phi_+, \Phi_-]^* = - I_{\mr{eff}}[\Phi_-,\Phi_+]\,,
	\qquad 
	\Im I_{\mr{eff}} \geq 0\,.
	\label{eq:sec2-SK-unitarity}
\end{align}
The first condition states that the influence of the environment disappears when the two SK histories coincide.
The second is the SK reality condition. 
The third condition makes the fluctuation kernel positive.

It is useful to introduce the $r$-$a$ variables by 
\begin{align}
	\Phi_r \coloneqq \frac{\Phi_+ + \Phi_-}{2}\,,
	\qquad 
	\Phi_a \coloneqq \Phi_+ - \Phi_-\,,
\end{align}
so that the response structure of the semiclassical SK action becomes manifest.
The average field $\Phi_r$ describes the phyiscal configuration and the coefficients of relative field $\Phi_a$ give the deterministic equations of motion (EOM) and retarded response in the linear term and the information of fluctuations and noises in the quadratic term.
Since the two copies coincide at the classical saddle, the semiclassical expansion is done around $\Phi_a = 0$.\footnote{
	Restoring $\hbar$, it is often useful to count the branch difference as $\Phi_+-\Phi_-=\hbar\,\Phi_a$, so that the linear $a$-term gives the classical saddle equation of $\exp(\iu I/\hbar)$.
	Although quadratic $a$-terms are then naively higher order, thermal noise can remain classical: by the fluctuation-dissipation relation, $\hbar\coth(\beta\hbar\omega/2)\to 2T/\omega$ at high temperature, leaving a finite noise kernel.} 
Up to quadratic order in $\Phi_a$, the system effective action takes the following form
\begin{align}
	I_{\mr{eff}}[\Phi_r, \Phi_a] = \int \dd^d x \, \Phi_a \msc{E}_r[\Phi_r(x)] + \frac{\iu}{2} \int \dd^dx \dd^d y \,\Phi_a(x) N(x,y;\Phi_r) \Phi_a(y) + \mc{O}(a^3)\,.
	\label{eq:sec2-ra-effective-action}
\end{align}
$\mc{O}(a^3)$ denotes the terms of the relative field cubed or greater.
We note that the $\mc{O}(a^0)$ term automatically vanishes due to the SK constraint \eqref{eq:sec2-SK-unitarity}.
The term linear in $\Phi_a$ gives the deterministic response equation for $\Phi_r$ denoted by $\msc{E}_r[\Phi_r]$, and the imaginary quadratic term gives the noise kernel.
The positivity condition in Eq.~\eqref{eq:sec2-SK-unitarity} implies that $N$ is positive as a kernel on the allowed fluctuation subspace.

The quadratic imaginary term admits an equivalent Hubbard--Stratonovich representation
\begin{align}
	&\exp \left[ - \frac{1}{2} \int \dd^dx \dd^dy \,\Phi_a (x) N(x,y;\Phi_r) \Phi_a(y) \right]
	\nn\\
	& \qquad 
	= \int \msc{D}\eta_\Phi \exp \left[
		- \frac{1}{2} \int \dd^dx \dd^d y \,\eta_\Phi(x) N^{-1}(x,y;\Phi_r) \eta_\Phi(y) - \iu \int \dd^dx \, \Phi_a(x) \eta_\Phi(x)
	\right]\,,
	\label{eq:sec2-HS}
\end{align}
where the overall normalization has been dropped and $N^{-1}$ is defined by 
\begin{align}
	\int \dd^dz \, N(x,z;\Phi_r) N^{-1}(z,y;\Phi_r) = \delta^d(x-y)\,.
\end{align}
The auxiliary field $\eta_\Phi$ is the induced Langevin noise.
We write the Gaussian average of $\msc{O}[\eta_\Phi]$ with fixed $\Phi_r$ as 
\begin{align}
	\braket{\msc{O}[\eta_\Phi]}_\eta \coloneqq \frac{\displaystyle \int \msc{D}\eta_\Phi \msc{O}[\eta_\Phi] \exp \left[ - \frac{1}{2} \int \dd^dx \dd^d y \,\eta_\Phi(x) N^{-1}(x,y;\Phi_r) \eta_\Phi(y) \right]}{\displaystyle \int \msc{D}\eta_\Phi \exp \left[ - \frac{1}{2} \int \dd^dx \dd^d y \, \eta_\Phi(x) N^{-1}(x,y;\Phi_r) \eta_\Phi(y) \right]}\,,
	\label{eq:sec2-noise-average}
\end{align}
then we obtain
\begin{align}
	\braket{\eta_\Phi(x)}_\eta=0\,,
	\qquad 
	\braket{\eta_\Phi(x) \eta_\Phi(y)}_\eta = N(x,y;\Phi_r)\,.
\end{align}
In this representation with the noise $\eta_\Phi$, the variation of $\Phi_a$ gives the stochastic EOM 
\begin{align}
	\msc{E}_r[\Phi_r(x)] = \eta_\Phi(x)\,.
\end{align}
Thus the deterministic saddle quation $\msc{E}_r[\Phi_r]=0$ is replaced with the Langevin equation with the noise $\eta_\Phi$ once the fluctuation is encoded in $\Im I_{\mr{eff}}$ explicitly. 
In thermal equilibrium, the dynamical Kubo--Martin--Schwinger (KMS) condition further relates the dissipative part of the response to the noise kernel (see Ref.~\cite{Liu:2018kfw}).

The system effective action may contain nonlocal kernel.
This nonlocality is already seen in Eq.~\eqref{eq:sec2-ra-effective-action}, where the noise kernel $N(x,y;\Phi_r)$ may correlate fields at different spacetime points.
When such kernels are nonlocal in the time direction, the system-only description becomes non-Markovian in the usual open-system sense because the response at a given time depends on the past history of the system. 
We refer to this history dependence as memory. 
In the following EFT construction, we keep the environmental response that mediates the exchange with the system as an explicit variable instead of starting from a nonlocal system-only kernel.
The environmental variable stores and relaxes the quantity exchanged with the system, and its retarded propagation becomes a memory kernel only after the variable is solved for with retarded boundary conditions and substituted back into the system-only description.

\paragraph{Simplified model for nonlocality from Langevin-type description.}

We demonstrate how a memory kernel arises after the retained environmental variable is solved for in a simple model.
As a minimal Langevin-type description of a non-conserved dissipative response, we introduce a retained environmental response variable $Y_r$ that couples to a system operator $\msc{O}_r$.
At the leading order in the derivative expansion, the environmental variable is assumed to obey the following stochastic relaxation equation 
\begin{align}
	(u\cdot \del + \Gamma) Y_r(x) = \chi \msc{O}_r(x) + \zeta_Y(x)\,.
	\label{eq:sec2-Markov-variable}
\end{align}
Here $\Gamma >0$ is the relaxation rate, $\chi$ is the response coefficient, and $\zeta_Y$ is the stochastic force for the retained environment variable.
In addition, we assume that $\zeta_Y$ is a local Gaussian noise satisfying
\begin{align}
	\braket{\zeta_Y(x)}_\zeta = 0\,,
	\qquad 
	\braket{\zeta_Y(x) \zeta_Y(y)}_\zeta = N_Y^{\mr{loc}}(x) \delta^d(x-y)\,,
\end{align}
where $\braket{\cdots}_\zeta$ denotes the stochastic average over $\zeta_Y$.
With the retarded boundary conditions, the retarted solution of \eqref{eq:sec2-Markov-variable} is expressed as 
\begin{align}
	Y_r(x) = \chi \int_0^\infty \dd s \, \ee^{-\Gamma s} \msc{O}_r(x- s u) + Y_{\mr{noise}}(x),
	\qquad 
	Y_{\mr{noise}}(x) = \int_0^\infty \dd s \,\ee^{-\Gamma s} \zeta_Y(x- s u)\,.
	\label{eq:sec2-eliminated-memory}
\end{align}
We note that $\ee^{-\Gamma s}$ is the retarded Green function of Eq.~\eqref{eq:sec2-Markov-variable}.
The nonlocality in this solution produces a memotry term depending on the past values of $\msc{O}_r$ along the environment flow when we integrate out $Y_r$ or substitute the solution \eqref{eq:sec2-eliminated-memory} into the system equations or action.
In particular, the two point function of the noise part in \eqref{eq:sec2-eliminated-memory} becomes 
\begin{align}
	\braket{Y_{\mr{noise}}(x) Y_{\mr{noise}}(y)}_\zeta = \int_0^\infty \dd s \dd s'\, \ee^{- \Gamma(s+s')} N^{\mr{loc}}_Y(x-su) \delta^d(x - su - y + s'u)\,,
	\label{eq:sec2-induced-colored-noise}
\end{align}
which is generally nonlocal as a function of $x$ and $y$.
Thus, the enlarged system-environment EFT provides the local and Markovian description in the variables $(\Phi, Y)$, but the coarse-grained system EFT obtained after integrating out the environment variable $Y$ contain retarded response and memory kernel.
In this sense, the enlarged system-environment EFT is regarded the Markovian embedding.

\subsection{Dissipative gravity as guiding example}
\label{subsec:SK-grav}

As the preparation for the non-Abelian open-EFT construction, we recall a lesson from the dissipative gravitational EFT discussed in Ref.~\cite{Lau:2024mqm}.
The dissipation in a coarse-grained system should be understood as the flow of conserved charges such as energy, momentum, or (gauge) charge, out of the variables kept in the system sector.
We introduce the retained environmental sector in the system-environment EFT so that we have non-vanishing dissipation between them and close the non-trivial Ward identity.
When we have a local symmetry, an $a$-type St\"uckelberg field is useful for writing the response of the charge current in a covariant form, but it is not by itself dynamical if we do not keep the environmental sector.
This is also the motivation for the introduction of the retained environmental sector in the system-environment EFT.
In the following parts of this section, we see these properties through reviewing the construction of the dissipative gravitational EFT.

Let us begin with the nondissipative situation for a scalar field theory in a curved spacetime.
We consider the following standard scalar field with the scalar potential $V(\phi)$:
\begin{align}
	S_\phi[\phi,g] = \int \dd^dx \sqrt{-g} \left[ -\frac{1}{2} g^{\mu\nu} \del_\mu \phi \del_\nu \phi - V(\phi) \right]\,.
\end{align}
The SK action becomes just the doubled conservative one as 
\begin{align}
	I_{\phi, \mr{cons}}[\phi_\pm, g_{\pm}] = S_\phi[\phi_+, g_+] - S_\phi[\phi_-,g_-]\,.
\end{align}
Expanding around the diagonal configuration with
\begin{align}
	\phi_\pm = \phi_r \pm \frac{\phi_a}{2}\,,
	\qquad 
	g_{\pm\mu\nu} = g_{r\mu\nu} \pm \frac{1}{2} g_{a\mu\nu}\,,
\end{align}
the doubled SK action is rewritten as 
\begin{align}
	I_{\phi,\mr{cons}}[\phi_{r,a},g_{r,a}] = \int \dd^dx \sqrt{-g_r} \left[
		\frac{1}{2} T^{\mu\nu}_{\phi_r} g_{a\mu\nu} + \left( \Box \phi_r - V'(\phi_r) \right) \phi_a
	\right] + \mc{O}(a^2)\,,
	\label{eq:sec2-cons-scalar-ra}
\end{align}
where $V'(\phi) = \dd V/ \dd \phi$ and $T^{\mu\nu}_{\phi}$ is the stress tensor for the scalar field $\phi$ in the $g_r$ background:
\begin{align}
	T^{\mu\nu}_\phi = \del^\mu\phi \del^\nu \phi - g^{\mu\nu} \left( \frac{1}{2} (\del\phi)^2 + V(\phi) \right)\,.
	\label{eq:scalar-em-tensor}
\end{align}
This is the ordinary closed-system SK doubling and no friction term is generated without the couplings with and the integration of additional environmental degrees of freedom.

As a naive system-only open-system model, we suppose that the high-energy modes have been integrated out and that their effect is kept only through the leading local term in the derivative expansion. This term is taken to be a friction term. The dissipative medium is not treated as a dynamical sector in this description, but is instead encoded in background data. Let $u^\mu$ be the local rest-frame vector of the dissipative medium and $\gamma_\phi$ a friction coefficient.
At the leading order of the relative fields, the effective action with the nondynamical medium can be written schematically as
\begin{align}
	I_{\phi,\mr{nondyn}}[\phi_{r,a}, g_{r,a}] = \int \dd^dx \sqrt{-g_r} \left[
		\frac{1}{2} T^{\mu\nu}_{\phi_r} g_{a\mu\nu} + \left( \Box \phi_r - \gamma_\phi u^\mu \del_\mu \phi_r - V'(\phi_r) \right) \phi_a 
	\right] + \mc{O}(a^2)\,.
	\label{eq:sec2-diss-scalar-bg}
\end{align}
The variation of $\phi_a$ gives the deterministic dissipative EOM 
\begin{align}
	\Box \phi_r - \gamma_\phi u^\mu \del_\mu \phi_r - V'(\phi_r) = 0\,.
	\label{eq:sec2-diss-scalar-eom}
\end{align}
The stress tensor in \eqref{eq:sec2-diss-scalar-bg} is still the conservative one in \eqref{eq:scalar-em-tensor} and its off-shell Ward identity is 
\begin{align}
	\nabla_\mu T^{\mu\nu}_{\phi_r} = (\Box \phi_r - V'(\phi_r)) \del^\nu \phi_r\,.
	\label{eq:sec2-cons-scalar-ward}
\end{align}
The dissipative EOM makes this become 
\begin{align}
	\nabla_\mu T^{\mu\nu}_{\phi_r} \stackrel{\mr{EOM}}{=} \gamma_\phi (u^\mu \del_\mu \phi_r) \del^\nu \phi_r\,.
	\label{eq:sec2-system-stress-nonconservation}
\end{align}
Here, $\nabla_\mu$ denotes the covariant derivative associated with $g_{r\mu\nu}$.
Thus the scalar sector is no longer separately conserved once it has a dissipative term.
This means that the additional stress tensor from the environment should be required to close the Ward identity for the nonvanishing $\gamma_\phi (u^\mu \del_\mu \phi_r) \del^\nu \phi_r$.
This right-handed side of \eqref{eq:sec2-system-stress-nonconservation} is the energy-momentum flux exchanged with the medium $u^\mu$ and $\gamma_\phi$.

The conservation breaking of the system conserved current can be understood from the viewpoint of the doubled local symmetry.
For the gravitational system, the nonconservation of the system stress tensor in Eq.~\eqref{eq:sec2-system-stress-nonconservation} is related to the doubled diffeomorphism symmetry.
In the effective action, the exchanged energy-momentum appears as  the obstruction to the relative or $a$-type diffeomorphism invariance.
The two SK branches have independent (infinitesimal) diffeomorphism with the parameters $\xi^\mu_\pm$.
On each branch, these act as 
\begin{align}
	\delta_\pm g_{\pm \mu\nu} = \pounds_{\xi_\pm} g_{\pm \mu\nu} = \nabla_\mu \xi_{\pm \nu} + \nabla_\nu \xi_{\pm \mu}\,,
	\qquad 
	\delta_{\pm} \phi_\pm = \pounds_{\xi_\pm} \phi_\pm = \xi^\mu_\pm \del_\mu \phi_\pm\,,
	\label{eq:sec2-branch-diffeomorphism}
\end{align}
where $\pounds_v$ denotes the Lie derivative along the vector field $v$.
We decompose the parameters in the $(r,a)$-basis as 
\begin{align}
	\xi_\pm^\mu \coloneqq \xi^\mu_r \pm \frac{1}{2} \xi^\mu_a\,.
\end{align}
The diagonal transformation $\delta_r$ generated by $\xi^\mu_r$ is the ordinary spacetime diffeomorphism acting on both the $r$-type fields and $a$-type fields.
The relative transformation $\delta_a$ is obtained at linear order by taking $\xi^\mu_\pm = \pm \xi^\mu_a/2$, which acts only on the $a$-type fields via the Lie derivative in the $r$-fields background.
Around a diagonal $r$-metric and at the leading order in the $a$-type fields, this $a$-type transformations are given by
\begin{align}
	\delta_a g_{a\mu\nu} = \nabla_\mu \xi_{a\nu} + \nabla_\nu \xi_{a\mu} = \pounds_{\xi_a} g_{r\mu\nu} \,,&
	\qquad 
	\delta_a \phi_a = \xi^\mu_a \del_\mu \phi_r
    = \pounds_{\xi_a} \phi_r\,,
	\\
	\delta_a g_{r\mu\nu} =0\,,&
	\qquad 
	\delta_a \phi_r = 0\,.
	\label{eq:sec2-noise-diff-transform}
\end{align}
For the conservative action \eqref{eq:sec2-cons-scalar-ra}, the $a$-type variation is proportional to the combination that appears in the conservative Ward identity:
\begin{align}
	\delta_a I_{\phi,\mr{cons}} = \int \dd^dx \sqrt{-g_r} \left[
		-\nabla_\mu T^{\mu\nu}_{\phi_r} + ( \Box \phi_r - V'(\phi_r)) \del^\nu \phi_r
	\right] \xi_{a\nu} + \mc{O}(a^2)\,,
\end{align}
Thus, the action is invariant under the $a$-type diffeomorphism when the Ward identity \eqref{eq:sec2-cons-scalar-ward} is imposed. Equivalently, requiring this invariance for  the $a$-type diffeomorphism transformation gives the Ward identity \eqref{eq:sec2-cons-scalar-ward}.
When we have the dissipation only in the system as Eq.~\eqref{eq:sec2-diss-scalar-bg}, this cancellation is spoiled by the friction term up to a boundary term as 
\begin{align}
	\delta_a I_{\phi,\mr{nondyn}} = - \int \dd^d x \sqrt{-g_r} \xi_{a\nu} \gamma_\phi (u^\mu \del_\mu \phi_r) \del^\nu \phi_r + \mc{O}(a^2)\,.
\end{align}
Thus, the system-only friction term is precisely the obstruction to the $a$-type diffeomorphism symmetry.
A locally imposed gauge symmetry is normally preserved pointwise, except at the boundary of the closed time path. While such a friction term may appear harmless from the viewpoint of $r$-type diffeomorphisms, it violates $a$-type diffeomorphism invariance. Therefore, even such an ordinary friction term is not allowed in a diffeomorphism invariant theory.\footnote{For gauge theories, a related top-down statement was established in Refs.~\cite{Kaplanek:2025moq,Kaplanek:2026kpp}: after BRST gauge fixing, the SK influence functional obtained by integrating out charged matter or hard modes retains the appropriate BRST/gauge constraints, with the closed-time-path boundary conditions treated explicitly.}
The natural first attempt is to compensate this transformation by introducing an $a$-type St\"uckelberg field.

We can therefore introduce an $a$-type St\"uckelberg field $X^\mu_a$ transformed as 
\begin{align}
	\delta_a X^\mu_a = \xi^\mu_a\,,
	\label{eq:sec2-Xa-noise-shift}
\end{align}
which is regarded an $a$-type environmental variable.
Let us define 
\begin{align}
	\mc{G}_{a\mu\nu} &\coloneqq g_{a\mu\nu} - \nabla_\mu X_{a\nu} - \nabla_\nu X_{a\mu}
    = g_{a\mu\nu} - \pounds_{X_a} g_{r\mu\nu}\,,
    \label{eq:sec2-grav-Stuckelberg-replacement-g}
	\\
	\varphi_a &\coloneqq \phi_a - X_a^\mu \del_\mu \phi_r
    = \phi_a - \pounds_{X_a} \phi_r\,,
	\label{eq:sec2-grav-Stuckelberg-replacement-phi}
\end{align}
so that they compensate the relative diffeomorphism.
If no $r$-type environmental variables are introduced, the St\"uckelberg-compensated action is given by 
\begin{align}
	I_\mr{naive} = \int \dd^d x \sqrt{-g_r} \left[
		\frac{1}{2} \left( - M_P^2 G_r^{\mu\nu} + T^{\mu\nu}_{\phi_r} \right) \mc{G}_{a\mu\nu} + \left( \Box \phi_r - \gamma_\phi u^\mu \del_\mu \phi_r - V'(\phi_r)  \right) \varphi_a
	\right] + \mc{O}(a^2)\,,
	\label{eq:sec2-naive-Stueck-gravity}
\end{align}
where $G_r^{\mu\nu} = R_r^{\mu\nu} - \frac{1}{2} R_r g_r^{\mu\nu}$ is the Einstein tensor for $g_{r\mu\nu}$ and $M_P$ is the reduced Planck scale.
After the partial integration, this action becomes 
\begin{align}
	I_{\mr{naive}} &= \int \dd^dx \sqrt{-g_r} \biggl[
		\frac{1}{2} \Bigl( - M_P^2 G^{\mu\nu}_r + T^{\mu\nu}_{\phi_r} \Bigr) g_{a\mu\nu} + (\Box \phi_r - \gamma_\phi u^\mu \del_\mu \phi_r - V'(\phi_r)) \phi_a
		\nn \\
		& \qquad \qquad \qquad \qquad 
		+ \gamma_\phi u^\mu \del_\mu \phi_r  \del_\nu \phi_r X^\nu_a
	\biggr] + \mc{O}(a^2)\,,
	\label{eq:sec2-naive-Stueck-gravity-expanded}
\end{align}
where the Einstein tensor and the scalar stress tensor satisfy
\begin{align}
	\nabla_\mu G^{\mu\nu}_r = 0\,,
	\qquad 
	\nabla_\mu T^{\mu\nu}_{\phi_r} = (\Box \phi_r - V'(\phi_r)) \del^\nu \phi_r\,.
\end{align}
The EOM of $X^\mu_a$ imposes 
\begin{align}
	\gamma_\phi u^\mu \del_\mu \phi_r \del_\nu \phi_r = 0\,.
	\label{eq:sec2-Xa-constraint}
\end{align}
Thus, the introduction of only the $a$-type St\"uckerlberg field or $a$-type environmental field removes the dissipative flux instead of providing symmetry preserving interactions with the environment.
This reason is that the rest-frame data entering the friction term such as $u^\mu$ and $\gamma_\phi$ remain nondynamical and we do not have the dynamical $r$-type fields in the environmental sector.

The open gravitational EFT avoids this problem by completing the $a$-type St\"uckelberg compensator into a genuine environmental $r$-$a$ pair.
In other words, $X_a^\mu$ is not introduced as an isolated field, but is accompanied by $r$-type environmental variables whose dynamics carries the exchanged energy and momentum.
The HydroEFT realization provides us a useful concrete example of this $r$-type environmental variables.
In the HydroEFT formulation\footnote{For details of the construction of HydroEFT, see Refs.~\cite{Crossley:2015evo,Glorioso:2016gsa,Glorioso:2017fpd,Liu:2018kfw}.}, these are maps from the physical spacetime to a material spacetime parameterized by 
\begin{align}
	\sigma^{\bar{m}}(x),\qquad
	\bar{m} = \bar{0},\ldots,\ol{d-1}\,.
\end{align}
Let $\ms{X}^\mu(\sigma)$ be the inverse embedding, the $r$-type environmental partner of the $a$-type St\"uckelberg field $X^\mu_a$.
Then, we define the pullback 
\begin{align}
	K^\mu{}_{\bar{m}}(x) \coloneqq \frac{\del \ms{X}^\mu(\sigma)}{\del \sigma^{\bar{m}}} \Bigg|_{\sigma = \sigma(x)}\,,
	\qquad 
	K^\mu{}_{\bar{m}} \del_\mu \sigma^{\bar{n}} = \delta^{\bar{n}}_{\bar{m}}\,.
\end{align}
The vector $K^\mu{}_{\bar{0}}$ is tangent to the environmental worldline at fixed material spatial labels.
Its unit vector becomes the local rest-frame vector
\begin{align}
	u^\mu = b^{-1} K^\mu{}_{\bar{0}},
	\qquad 
	b \coloneqq \sqrt{-g_{\mu\nu} K^\mu{}_{\bar{0}} K^\nu{}_{\bar{0}}}\,.
\end{align}
The important change here is the status of $u^\mu$.
In the system-only description, $u^\mu$ was treated as a fixed rest-frame vector as discussed for \eqref{eq:sec2-diss-scalar-bg} and \eqref{eq:sec2-naive-Stueck-gravity}.
In the local system-environment EFT discussed here, it is built from the retained environmental maps, which is the $r$-type variables in the environment sector.

In the following discussion, we mainly use HydroEFT as a way to keep environmental variables explicit rather than relying on special properties of hydrodynamic modes.
What is required in dynamical gravity is an environmental stress tensor that closes the Ward identity. 
HydroEFT supplies such a tensor when the environment admits a derivative expansion:
\begin{align}
	T^{\mu\nu}_{\mr{env}} = \rho(\beta) u^\mu u^\nu + p(\beta) \Delta^{\mu\nu}_u + \cdots,
	\qquad 
	\Delta^{\mu\nu}_u = g^{\mu\nu} + u^\mu u^\nu\,.
	\label{eq:sec2-env-hydro-stress}
\end{align}
where the ellipsis denotes derivative corrections and possible additional environmental fields.
Here, $\beta$ is the local inverse temperature with $T\coloneqq \beta^{-1}$, and the equation of state is specified by $\rho=\rho(\beta)$ and $p=p(\beta)$.

Now, we find the system-environment total action for dissipative gravitational system.
At the linear order in the relative fields, the total SK action is given schematically by 
\begin{align}
	I_{\mr{tot}} &= \int \dd^dx \sqrt{-g_r} \biggl[
		\frac{1}{2} \Bigl( - M_P^2 G^{\mu\nu}_r + T^{\mu\nu}_{\phi_r} + T^{\mu\nu}_{\mr{env}} \Bigr) \mc{G}_{a\mu\nu} + (\Box \phi_r - \gamma_\phi u^\mu \del_\mu \phi_r - V'(\phi_r)) \varphi_a
	\biggr] + \mc{O}(a^2)
	\label{eq:sec2-total-open-grav-action}
	\\
	&= \int \dd^dx \sqrt{-g_r} \Biggl[
		\frac{1}{2} \Bigl( - M_P^2 G^{\mu\nu}_r + T^{\mu\nu}_{\phi_r} + T^{\mu\nu}_{\mr{env}} \Bigr) g_{a\mu\nu} + (\Box \phi_r - \gamma_\phi u^\mu \del_\mu \phi_r - V'(\phi_r)) \phi_a
		\nn \\
		&\qquad 
		+ \bigg\{
			 \nabla_\mu \Bigl( - M_P^2 G^{\mu\nu}_r + T^{\mu\nu}_{\phi_r} + T^{\mu\nu}_{\mr{env}} \Bigr) - (\Box \phi_r - \gamma_\phi u^\mu \del_\mu \phi_r - V'(\phi_r))\del^\nu \phi_r 
		\biggr\} X_{a\nu}
	\Biggr] + \mc{O}(a^2)
	\label{eq:sec2-total-open-grav-action-2}
\end{align}
The variation with respect to $X^\mu_a$ now gives the exchange equation for the environmental stress tensor 
\begin{align}
	\nabla_\mu T^{\mu\nu}_{\mr{env}} = - \gamma_\phi u^\mu\del_\mu\phi_r \del^\nu \phi_r + \cdots\,.
	\label{eq:sec2-env-stress-exchange}
\end{align}
In addition, \eqref{eq:sec2-system-stress-nonconservation} and \eqref{eq:sec2-env-stress-exchange} give the Ward identity 
\begin{align}
	\nabla_\mu \left( T^{\mu\nu}_{\phi_r} + T^{\mu\nu}_{\mr{env}} \right) = 0\,.
	\label{eq:sec2-total-Ward-id-stress-tensor}
\end{align}
The metric couples to the both stress tensors and the Einstein equation given by the variation of $g_{a\mu\nu}$ becomes 
\begin{align}
	M_P^2 G^{\mu\nu}_r = T^{\mu\nu}_{\phi_r} + T^{\mu\nu}_{\mr{env}}\,,
\end{align}
as required by the universal coupling of gravity.

Before summarizing the lesson from the gravitational system, we comment on higher-order terms in the $a$-fields. The terms linear in the $a$-fields control the deterministic equations and the associated Ward identities, whereas the quadratic $a$-field terms encode noise kernels. Such terms are especially important for thermal fluctuations, which can contribute already at quadratic order and are tied to dissipation by fluctuation-dissipation or KMS relations. We do not review these noise terms here, since the purpose of this section is to explain the relation among dissipation, symmetry, and conservation laws, and to identify the Ward-complete local system-environment structure.

\paragraph{Lesson of this section.}

The lesson from the gravitational construction is that an environmental sector is natural and useful in a bottom-up EFT for open systems with long-range forces. Once environmental variables are integrated out, the system-only response is generically encoded in nonlocal kernels, whose symmetry-constrained structure can be complicated. A bottom-up local EFT avoids taking this full nonlocal kernel as an input from the outset. Instead, it keeps the relevant environmental variables explicitly, so that the exchanged energy and momentum are carried locally and the spacetime Ward identity closes before the environmental variables are integrated out.

The non-Abelian construction adopts the same bottom-up viewpoint. The gauge covariance can be maintained by introducing a St\"uckelberg-type color-frame variable, but this symmetry completion only tells us how the environmental sector transforms. It does not by itself provide the current that carries the color charge exchanged with the YM field. Therefore, a local system-environment EFT should include an environmental color-current sector. This sector carries the exchanged color charge and allows the covariant Ward identity to close for the total color current of the colored degrees of freedom. Nonlocal system-only kernels appear only after the retained environmental variables are integrated out.

\section{Open EFT for non-Abelian gauge theory}
\label{sec:NA-gauge}

We now construct the non-Abelian analogue of the local system-environment EFT reviewed in Sec.~\ref{subsec:SK-grav}, where the system field is the YM field and color charged matter fields.
As in the gravitational construction, a local dissipative EFT should keep track of the environmental sector that carries the conserved quantity exchanged with the system.
In the YM case, this conserved quantity is color charge and the local covariant Ward identity closes when the environmental contribution to the color current is included together with the system contribution. 
For the $r$-$a$ pair of the environmental sector as $\ms{X}^\mu(\sigma)$ and $X^\mu_a$ in the gravitational case, we consider group valued field $h_r$ and the $a$-type St\"uckelberg field $\pi_a$, which gives the gauge-covariant description of the environmental sector and interaction between the system variables and environmental color current.
Nonlocal dissipative kernels arise only after the retained environmental variables are solved with retarded boundary conditions and substituted back into the system-only description. 
Throughout this section, $u^\mu$ denotes the environmental velocity introduced in Sec.~\ref{sec:SK-formalism}.

The goal of this section to give the local color-sector embedding explicitly.
Sec.~\ref{sec:relative-gauge-frame} fixes the non-Abelian conventions and constructs the environmental color frame field $h_r$ and $a$-type partner $\pi_a$.
Sec.~\ref{sec:color-fluid-completion} makes this sector dynamical by adding an environmental color current with local constitutive response, so that the total color Ward identity closes locally. 
Sec.~\ref{sec:color-frame-env-elimination} formulates a general template for integrating out retained environmental response variables and shows how it produces nonlocal system-only kernels, and Sec.~\ref{sec:gauge-invariant-fermions} gives a matter-channel illustration before the hard-loop realization in Sec.~\ref{sec:HTL-matching}.
In the following sections, we work in flat spacetime only to simplify the discussion and keep the focus on the color and SK structures. 
The extension to a curved SK background is straightforward by coupling the SK metric sources to the system-environment EFT for the non-Abelian gauge theory, where the Ward identities are given in the covariant forms.

\subsection{SK gauge covariance and environmental variables}
\label{sec:relative-gauge-frame}

This subsection fixes the variables used in the following discussion and derive the effective action for the dissipative YM theory.
We first state the non-Abelian gauge theory conventions, then introduce the doubled SK gauge fields and $r$-$a$ type gauge transformations.
We introduce the environmental color frame field $h_r$ and the $a$-type St\"uckelberg field $\pi_a$, and discuss the interactions between the system and the environment and the Ward identity.
To clarify the discussion and calculations in this paper, we consider the gauge theory with the gauge group being $G = \SU(N)$ in mind.
The hermitian generators $T^A$ $(A =1,\ldots, \dim G)$ are normalized as 
\begin{align}
	\Tr T^A T^B = \frac{1}{2} \delta^{AB}\,.
\end{align}
The local construction only uses this trace normalization and the adjoint transformation law, and can be adapted to other compact gauge groups by replacing the trace normalization and the representation matrices accordingly.

We introduce the covariant derivative associated with the gauge field or gauge connection $A_\mu \in \mf{g} = \mf{su}(N)$ by 
\begin{align}
	D_\mu [A] \coloneqq \del_\mu + \iu \gc A_\mu\,,
	\label{eq:sec3-D-fund}
\end{align}
where $\gc$ is the gauge coupling constant.
For the gauge transformation by $U \in G$, the covariant derivative transforms covariantly as 
\begin{align}
	D_\mu[A^U] = U D_\mu[A] U^{-1}\,,
\end{align}
where $A^U_\mu$ denotes the transformed gauge field of $A_\mu$ with $U$.
This leads to 
\begin{align}
	A_\mu \mapsto A^U_\mu = UA_\mu U^{-1} - \frac{\iu}{\gc} U \del_\mu U^{-1}\,.
	\label{eq:sec3-gauge-transform}
\end{align}
The commutator of this covariant derivative gives the field strength 
\begin{align}
	[D_\mu[A], D_\nu[A]] = \iu \gc F_{\mu\nu}\,,
\end{align}
namely 
\begin{align}
	F_{\mu\nu} = \del_\mu A_\nu - \del_\nu A_\mu + \iu \gc [A_\mu, A_\nu]\,.
	\label{eq:sec3-F}
\end{align}
This curvature transforms covariantly under the gauge transformation by $U$ as 
\begin{align}
	F_{\mu\nu} \mapsto U F_{\mu\nu} U^{-1}\,,
\end{align}
which becomes gauge invariant in the Abelian case.
In the current normalization, the YM action is given by 
\begin{align}
	S_\YM[A] = - \frac{1}{2} \int \dd^dx \Tr F_{\mu\nu} F^{\mu\nu}\,,  
	\label{eq:se3-YMaction}
\end{align}
and the EOM for the pure YM gauge field is 
\begin{align}
	D_\mu[A] F^{\mu\nu} = \del_\mu F^{\mu\nu} + \iu \gc [A_\mu, F^{\mu\nu}] = 0\,.
\end{align}

Let us introduce the $r$-$a$ basis for the doubled SK gauge fields $A_{\pm \mu}$ in the same manner with Sec.~\ref{sec:SK-formalism} by 
\begin{align}
	A_{\pm \mu} = A_{r\mu} \pm \frac{1}{2} A_{a\mu}\,.
	\label{eq:sec3-A-ra}
\end{align}
The doubled SK gauge fields can be transformed independently on each branch 
\begin{align}
	A_{\pm \mu} \mapsto A^{U_\pm}_{\pm \mu} = U_\pm A_{\pm \mu} U_\pm^{-1} - \frac{\iu}{\gc} U_\pm \del_\mu U_\pm^{-1}\,,
\end{align}
where $U_\pm$ denotes the doubled gauge transformation matrix on the SK branch.
The diagonal $r$-type transformation is given by setting $U_+ = U_- = U_r$ and the gauge fields in $r$-$a$ basis transform as
\begin{align}
	A_{r\mu} \mapsto A^{U_r}_{r\mu} = U_r A_{r\mu} U_r^{-1} - \frac{\iu}{\gc} U_r \del_\mu U_r^{-1}\,,
	\qquad 
	A_{a\mu} \mapsto U_r A_{a\mu} U_r^{-1}\,.
	\label{eq:sec3-Ara-transform}
\end{align}
It is found that $A_{r\mu}$ transforms as a usual gauge connection and $A_{a\mu}$  enjoys the homogeneous transformation.

Next, let us consider the decomposition of the doubled gauge transformation near the diagonal subgroup by expanding the relative gauge group element.
Because the gauge transformation matrix is group valued, there are some ambiguities of the parametrization unlike the system fields.
Here, we parametrize the doubled gauge transformation matrix in the factorized form as 
\begin{align}
	U_\pm = U_r \exp \left( \pm \frac{\iu}{2} \alpha_a \right)\,,
	\qquad 
	\alpha_a \in \mf{g}\,,
	\label{eq:sec3-U-ra}
\end{align}
which leads to 
\begin{align}
	U_-^{-1} U_+ = \exp(\iu \alpha_a)\,.
	\label{eq:sec3-U-relative}
\end{align}
In this sense, $\exp(\iu \alpha_a)$ parametrizes the relative perturbation around the diagonal $r$-type element $U_r$.
The parameter $\alpha_a \in \mf{g}$ is the relative or $a$-type gauge parameter in this product coordinate, which gives a local Lie-algebra coordinate for $U_-^{-1}U_+$ near the identity.\footnote{
This $\alpha_a$ introduces a local coordinate on the group manifold.
The choice of \eqref{eq:sec3-U-ra} or \eqref{eq:sec3-U-relative} is smooth and unambiguous only within a chosen logarithmic patch.
When we consider the large gauge transformations or global structure on the group manifold, we should keep the finite relative group element $U_-^{-1}U_+$ rather than the parametrization of $\alpha_a$.
On the other hand, this issue does not arise in the present paper, since we work within the semiclassical regime.}
If one locally writes $U_\pm = \exp (\iu \alpha_\pm)$ and $U_r = \exp(\iu \alpha_r)$, Eq.~\eqref{eq:sec3-U-ra} becomes 
\begin{align}
	U_\pm = \exp(\iu \alpha_\pm) = \exp(\iu \alpha_r) \exp\left( \pm \frac{\iu}{2} \alpha_a \right)\,.
\end{align}
At the leading order of $\alpha$, this reduces to 
\begin{align}
	\alpha_\pm = \alpha_r \pm \frac{1}{2} \alpha_a\,,
\end{align}
which is the same $r$-$a$ parametrization with the system field including the gauge field in this section.
We note that the parametrization \eqref{eq:sec3-U-ra} is not identified with the additive branch-coordinate parametrization 
\begin{align}
	U_\pm = \exp \left[ \iu \left( \alpha_r \pm \frac{1}{2} \tilde{\alpha}_a \right) \right]\,,
	\label{eq:sec3-U-ra-2}
\end{align}
except the leading order of $\alpha$'s due to the non-vanishing commutator terms.
The parametrization of \eqref{eq:sec3-U-ra} and \eqref{eq:sec3-U-ra-2} coincides in the Abelian case.
In the nonlinear representation, the relative variables should be parameterized by the factorized form as in Eq.~\eqref{eq:sec3-U-ra} instead of by the difference of the Lie algebra variables.

A pure $a$-type transformation is obtained by taking $U_r = \bm{1}$ in \eqref{eq:sec3-U-ra}
\begin{align}
	U_\pm = \exp \left( \pm \frac{\iu}{2} \alpha_a \right)\,.
	\label{eq:sec3-a-gauge-U}
\end{align}
For a finite $a$-type transformation, the both $r$- and $a$-type variables transform nonlinearly.
In the semiclassical SK expansion, however, we are interested in infinitesimal transformation at leading order in the $a$-fields and focus on the leading order forms of $a$-type variables.
The infinitesimal gauge transformation in the branch-basis is 
\begin{align}
	\delta_\pm A_{\pm \mu} = - \frac{1}{\gc} D_\mu[A_\pm] \alpha_\pm\,,
\end{align}
then we find that 
\begin{align}
	\delta_a A_{r\mu} &= \frac{1}{2} (\delta_+ A_{+ \mu} + \delta_- A_{-\mu})
	= - \frac{\iu}{4} [A_{a\mu}, \alpha_a] = \mc{O}(a^2)\,,
	\\
	\delta_a A_{a\mu} &= \delta_+ A_{+\mu} - \delta_- A_{-\mu} = - \frac{1}{\gc} D_\mu[A_r] \alpha_a\,.
	\label{eq:sec3-a-gauge-Aa}
\end{align}
At the leading order in the $a$-counting, the pure $a$-type gauge transformation leaves the diagonal field unchanged and acts on the $a$-type gauge field inhomogeneously
\begin{align}
	\delta_a A_{r\mu} = 0\, ,
	\qquad 
	\delta_a A_{a\mu} = - \frac{1}{\gc} D_\mu[A_r] \alpha_a\,.
	\label{eq:sec3-a-gauge-leading}
\end{align}

\subsubsection*{Introduction of environmental color frame field and St\"uckelberg field}

Now we introduce the environmental color frame in order to couple the environment to the system.
For clarity of the gauge-invariant construction, we introduce the $r$-type fields first, in the order opposite to the gravitational case, and then introduce the $a$-type fields by analogy with gauge transformations.
The retained $r$-type frame field $h_r(x)$ is a group valued environmental variable transforming as 
\begin{align}
	h_r(x) \in G\,,
	\qquad 
	h_r \mapsto U_r h_r\,,
	\label{eq:sec3-h-transform}
\end{align}
under the left-action of the diagonal gauge transformation $U_r(x) \in G$.
This $h_r$ fixes a local color basis for the environmental sector, then we call $h_r$ the environmental color frame field or simply the color frame field.
In this frame, environmental charges, currents, and response variables are written as local matrix-valued quantities.
This frame field introduces the dressed connection by\footnote{
It is noted that this dressed connection $\tilde{\mc{A}}_{r\mu}$ is define formally just in the $r$-basis, which does not come from the field redefinition as $\mc{A}_{\pm \mu} \to \mc{A}_{ra\mu}$.
The relations between this $\tilde{\mc{A}}_{r\mu}$ and $\mc{A}_{r\mu}$ is given by \eqref{eq:sec3-calA-r-expand} and they coincide up to the $\mc{O}(a^2)$ order.
}
\begin{align}
	\tilde{\mc{A}}_{r\mu} \coloneqq h_r^{-1} A_{r\mu} h_r - \frac{\iu}{\gc} h_r^{-1} \del_\mu h_r\,.
	\label{eq:sec3-caltildeA-def}
\end{align}
This dressed field is invariant under the diagonal transformation 
\begin{align}
	\tilde{\mc{A}}_{r\mu}^{U_r} = \tilde{\mc{A}}_{r\mu}\,,
	\label{eq:sec3-caltildeA-invariant}
\end{align}
and represents the soft gauge field in the environmental color frame.
This $h_r$ is also called the dressing field because its coupling produces the gauge invariant dressed field.
The corresponding dressed field strength and covariant derivative are defined by
\begin{align}
	\tilde{\mc{F}}_{r\mu\nu} &\coloneqq \del_\mu \tilde{\mc{A}}_{r\nu} - \del_\nu \tilde{\mc{A}}_{r\mu} + \iu \gc [\tilde{\mc{A}}_{r\mu}, \tilde{\mc{A}}_{r\nu}] 
	= h_r^{-1}F_{r\mu\nu}h_r\,,
	\label{eq:sec3-caltildeF-def}
	\\
	\mc{D}_\mu[\tilde{\mc{A}}_r] &\coloneqq \del_\mu + \iu \gc \tilde{\mc{A}}_{r\mu}\,,
	\label{eq:sec3-calD-def}
\end{align}
where $F_{r\mu\nu} \coloneqq \del_\mu A_{r\nu} - \del_\nu A_{r\mu} + \iu \gc [A_{r\mu}, A_{r\nu}]$.
It is noted that this covariant derivative satisfies 
\begin{align}
    \mc{D}_\mu [\tilde{\mc{A}}_r] = h_r^{-1} D_\mu[A_r] h_r.
\end{align}
$h_r$ gives the interactions between the system and the environment, and the dissipative dynamics will be encoded in the constitutive current of the dynamical $h_r$-sector as discussed in the following subsection.

Having introduced $h_r$ to form the diagonal gauge invariant $r$-type variables, we now lift the construction to the doubled SK framework using the right-trivialization as \eqref{eq:sec3-U-ra}.
Near the diagonal configuration $h_+ = h_- = h_r$, we introduce the $a$-type St\"uckelberg field by the factorized form as
\begin{align}
	h_\pm \eqqcolon h_r \exp \left( \pm \frac{\iu}{2} \pi_a \right)
	\approx h_r \left( \bm{1} \pm \frac{\iu}{2} \pi_a \right)\,,
	\qquad 
	\pi_a \in \mf{g}\,,
\end{align}
which leads to 
\begin{align}
	h_-^{-1} h_+ = \exp (\iu \pi_a)\,.
\end{align}
This $\pi_a$ is the right-trivialized relative color frame variable related to the relative gauge parameter $\alpha_a$ in \eqref{eq:sec3-U-ra} and compensates the relative SK gauge transformation as same with $X^\mu_a$ for the relative diffeomorphism.
Under the relative gauge transformation, this St\"uckelberg field transforms as 
\begin{align}
	\delta_a \pi_a = h_r^{-1} \alpha_a h_r + \mc{O}(a^2)\,.
	\label{eq:sec3-a-gauge-pia}
\end{align}

Let us define the dressed doubled and the dressed $r$-$a$ basis gauge fields by 
\begin{equation}
	\mc{A}_{\pm \mu} \coloneqq h^{-1}_\pm A_{\pm \mu} h_\pm - \frac{\iu}{\gc} h_\pm^{-1} \del_\mu h_\pm\,,
\end{equation}
\begin{equation}
	\mc{A}_{r\mu} \coloneqq \frac{1}{2} (\mc{A}_{+\mu} + \mc{A}_{-\mu})\,,
	\qquad 
	\mc{A}_{a\mu} \coloneqq \mc{A}_{+\mu} - \mc{A}_{-\mu}\,.
	\label{eq:sec3-calA-ra}
\end{equation}
Expanding the $a$-fields in \eqref{eq:sec3-calA-ra} up to $\mc{O}(a^2)$, we find that the following expressions for the dressed $r$-$a$ basis gauge fields 
\begin{align}
	\mc{A}_{r\mu} &= h_r^{-1} A_{r\mu} h_r - \frac{\iu}{\gc} h_r^{-1} \del_\mu h_r + \mc{O}(a^2) = \tilde{\mc{A}}_{r\mu} + \mc{O}(a^2)\,,
	\label{eq:sec3-calA-r-expand}
	\\
	\mc{A}_{a\mu} &= h_r^{-1} A_{a\mu} h_r + \frac{1}{\gc} \mc{D}_\mu[\mc{A}_r] \pi_a + \mc{O}(a^2)\,.
	\label{eq:sec3-calA-a-expand}
\end{align}
We note that $\mc{A}_{r\mu}$ agrees with \eqref{eq:sec3-caltildeA-def} up to $\mc{O}(a^2)$.
$\mc{A}_{a\mu}$ in \eqref{eq:sec3-calA-a-expand} is the diagonal gauge invariant $a$-type gauge field built from $A_{a\mu}$ and the $a$-type St\"uckelberg field $\pi_a$ as similar to $\mc{G}_{a\mu\nu}$ with $g_{a\mu\nu}$ and $X^\mu_a$ in the gravitational system.
It is found that $\delta_a \mc{A}_{a\mu} = \mc{O}(a^2)$ due to Eq.~\eqref{eq:sec3-a-gauge-pia}.
We write the field strength calculated by $\mc{A}_{r\mu}$ in \eqref{eq:sec3-caltildeF-def} as $\mc{F}_{r\mu\nu}$.

\paragraph{Case of Abelian gauge theory: $G = \U(1)$.}

When we consider the Abelian gauge theory with $G= \U(1)$, the all commutators vanish and we find that 
\begin{align}
	h_\pm = h_r \exp \left( \pm \frac{\iu}{2} \pi_a \right)
	= \exp (\iu \pi_r) \exp \left( \pm \frac{\iu}{2} \pi_a \right)
	= \exp \left[
		\iu \left( \pi_r \pm \frac{1}{2} \pi_a \right)
	\right]\,,
\end{align}
where $\pi_r, \pi_a \in \mbb{R}$.
In this case, the $r$-type color frame field and the $a$-type St\"uckelberg field enjoy the same additive relations to the doubled SK fields with the system fields.
The dressed gauge fields in the $r$-$a$ basis are 
\begin{align}
    \mc{A}_{r\mu} = A_{r\mu} + \frac{1}{\gc} \del_\mu \pi_r,
    \qquad 
    \mc{A}_{a\mu} = A_{a\mu} + \frac{1}{\gc} \del_\mu \pi_a\,.
    \label{eq:sec3-abelian-limit-color-frame}
\end{align}
$\pi_{r,a}$ is the usual (Abelian) St\"uckelberg scalar field and the dressed gauge field squared term has the same form with the St\"uckelberg coupling
\begin{align}
	\mc{A}_\mu^2 =  \left( A_\mu + \frac{1}{\gc} \del_\mu \pi \right)^2\,,
\end{align}
which is invariant under the gauge transformation of $A_\mu$ and shift of $\pi$ as 
\begin{align}
	A_\mu \mapsto A_\mu - \frac{1}{\gc} \del_\mu \theta\,,
	\qquad 
	\pi \mapsto \pi + \theta\,,
\end{align}
with the gauge transformation parameter $\theta \in \mbb{R}$.

\subsubsection*{Effective action for dissipative gauge sector}

Now, let us derive the effective action for the dissipative gauge theory.
First, we begin with the ordinary doubled YM theory without the dissipation, which action is 
\begin{align}
	I^{(0)}_{\YM} = S_\YM[A_+] - S_\YM[A_-]\,,
	\label{eq:sec3-pure-ym-sk}
\end{align}
with the YM action~\eqref{eq:se3-YMaction}.
This system action is gauge invariant on each SK branch and does not require the environmental frame variables $h_r$ or $\pi_a$ due to no dissipation.
The expansion around the diagonal configuration gives 
\begin{align}
	I^{(0)}_\YM [A_r,A_a]= 2 \int \dd^dx \Tr A_{a\mu} D_\nu[A_r] F^{\nu\mu}_r + \mc{O}(a^2)\,.
	\label{eq:sec3-pure-ym-linear}
\end{align}
Thus, in the absence of an environment, the variation of $A_{a\mu}$ imposes the classical YM equation $D_\mu[A_r]F_r^{\mu\nu}  =0$ for $A_{r\mu}$.
The identity coming from the relative gauge transformation of \eqref{eq:sec3-pure-ym-linear}
\begin{align}
	D_{\mu}[A_r] D_\nu[A_r] F^{\nu \mu }_r=0,
	\label{eq:sec3-ym-noether-identity}
\end{align}
is the Noether identity associated with the diagonal gauge redundancy of the pure YM action.

In order to couple the system with the environment and have a dissipation, we introduce the environmental color frame field $h_r$ and the $a$-type St\"uckelberg field $\pi_a$ and dress the system field in these frame fields.
The effective action becomes
\begin{align}
	I_{\YM} = 2 \int \dd^dx \Tr \mc{A}_{a\mu} \mc{D}_\nu[\mc{A}_r] \mc{F}_r^{\nu\mu} + I_{hJ} + \mc{O}(a^2)\,,
	\label{eq:sec3-YA-effective-action}
\end{align}
and the gauge field-current interaction term is assumed to have the following form:
\begin{align}
	I_{hJ} = - 2 \int \dd^dx \Tr \mc{A}_{a\mu} (\mc{J}^\mu_h + \mc{J}^\mu_\sys) + \mc{O}(a^2)\,.
	\label{eq:sec3-calA-current-coupling}
\end{align}
$\mc{J}^\mu_h$ is the environmental color current in the frame characterized by $h_r$.
We often call this current as the environmental $h_r$ current or $h_r$ current simply.
This is the color current analogy of $T^{\mu\nu}_{\mr{env}}$ in the gravitational theory.
$\mc{J}^\mu_\sys$ denotes the color current carried by the system charged field, which corresponds to $T^{\mu\nu}_{\phi}$ in the gravitational theory, but we do not specify the matter field properties here.
The total action becomes 
\begin{align}
	I_\YM &=2 \int \dd^dx \Tr \mc{A}_{a\mu} \Bigl(
		\mc{D}_\nu [\mc{A}_r] \mc{F}^{\nu\mu}_r - (\mc{J}^\mu_h + \mc{J}^\mu_\sys)
	\Bigr)  + \mc{O}(a^2)
	\\
	&= 2 \int \dd^dx \Tr A_{a\mu} \Bigl( D_\nu[A_r] F^{\nu\mu}_r - J^\mu_h - J^\mu_\sys \Bigr)
	\nn\\
	& \qquad
	- \frac{2}{\gc} \int \dd^dx \Tr \pi_a \mc{D}_\mu[\mc{A}_r] \Bigl(
		\mc{D}_\nu[\mc{A}_r] \mc{F}^{\nu\mu}_r - \mc{J}^\mu_h - \mc{J}^\mu_\sys
	\Bigr) + \mc{O}(a^2)\,
	\label{eq:sec3-total-effective-action-YM}
\end{align}
with 
\begin{align}
	J^\mu_h \coloneqq h_r \mc{J}^\mu_h h^{-1}_r\,,
	\qquad 
	J^\mu_\sys \coloneqq h_r \mc{J}^\mu_\sys h_r^{-1}\,.
	\label{eq:sec3-physical-env-current}
\end{align}
The variation of $A_{a\mu}$ provides the YM equation with the color current 
\begin{align}
	D_\mu[A_r] F_r^{\mu\nu} - J^\nu_h - J^\nu_\sys =0\,.
	\label{eq:sec3-sourced-ym-eom}
\end{align}
The second term in \eqref{eq:sec3-total-effective-action-YM} proportional to $\pi_a$ produces the covariant Ward identity including the environmental current
\begin{align}
	\mc{D}_\mu[\mc{A}_r] \left( \mc{J}^\mu_h + \mc{J}^\mu_\sys \right) = h_r^{-1} D_\mu[A_r] \left( J^\mu_h + J^\mu_\sys \right) h_r = 0\,.
	\label{eq:sec3-h-current-ward}
\end{align}
This condition is the current-sector Ward equation associated with the same gauge redundancy that gives the Noether identity \eqref{eq:sec3-ym-noether-identity} for the pure YM system.

\paragraph{Dissipation in gauge sector and Ohmic flux.}

For the gauge sector, the electric field $E_r^\mu \coloneqq F_r^{\mu\nu} u_\nu$ in the gauge field EOM plays the role of the friction term as $u^\mu \del_\mu \phi_r$ for the scalar field discussed in Sec.~\ref{subsec:SK-grav}.
In the current EOM \eqref{eq:sec3-sourced-ym-eom} with $J^\mu_\sys = 0$, the transverse Ohmic flux gives the simplest example
\begin{align} 
	J^\mu_{\mr{Ohm}} = \sigma_h E^\mu_r = \sigma_h F_r^{\mu\nu} u_\nu\,.
\end{align}
$\sigma_h$ is the color conductivity.
In this case, the effective action becomes 
\begin{align}
	I_{\YM,\mr{Ohm}} &=2 \int \dd^dx \Tr A_{a\mu} \Bigl( D_\nu[A_r] F^{\nu\mu}_r - J^\mu_{\mr{Ohm}}\Bigr)
	\nn\\
	& \qquad
	- \frac{2}{\gc} \int \dd^dx \Tr \pi_a \mc{D}_\mu[\mc{A}_r] \Bigl(
		\mc{D}_\nu[\mc{A}_r] \mc{F}^{\nu\mu}_r - \mc{J}^\mu_{\mr{Ohm}}
	\Bigr) + \mc{O}(a^2)\,
	\label{eq:sec2-naive-Stueck-gauge}
\end{align}
However, this current divergence vanishes because of the EOM and the Ward identity:
\begin{align}
	\mc{D}_\mu [\mc{A}_r] \mc{J}^\mu_{\mr{Ohm}} = \mc{D}_\mu[\mc{A}_r]\left( \sigma_h \mc{E}^\mu_r \right)
	= h_r^{-1} D_\mu[A_r] \left( \sigma_h E^\mu\right) h_r 
	=0\,,
\end{align}
and we do not have dissipations from the system.
Here, we introduce the dressed electric field and use the following relation:
\begin{align}
	\mc{E}^\mu_r \coloneqq \mc{F}^{\mu\nu}_r u_\nu\,,
	\qquad 
	\mc{E}^\mu_r = h_r^{-1} E^\mu_r h_r\,.
\end{align}

This is the analogy for the gravitational system described by \eqref{eq:sec2-naive-Stueck-gravity}.
In the both cases of \eqref{eq:sec2-naive-Stueck-gravity} and \eqref{eq:sec2-naive-Stueck-gauge}, we have apparent friction terms but no dissipation.
As in Sec.~\ref{subsec:SK-grav}, we introduce the environmental contributions to provide the nontrivial dissipation and interactions between the system and the environment in the next subsection.

\subsection{Local color-frame environment and color dissipation}
\label{sec:color-fluid-completion}

We now construct the local color frame environmental sector in the system-environment EFT and give the explicit form to the $h_r$ current, which accounts for the color charge exchanged through the dissipation and closes the total color Ward identity.
In analogy with the gravitational example, the $r$-type environmental variables carry the exchanged color charge and the $a$-type environmental field imposes the color Ward identity including the environment contributions.
The hydrodynamic variables introduced in Sec.~\ref{subsec:SK-grav}, such as $u^\mu$, $T$, $\Delta_u^{\mu\nu}$, and the environment stress tensor $T^{\mu\nu}_{\env}$ remain part of the environmental sector but we use these only through the local color constitutive relations in this subsection.

The color sector constitutive relations are written in the environmental color frame.
We define the electric field and the chemical potential in the color frame by 
\begin{align}
	\mc{E}^\mu_r \coloneqq \mc{F}^{\mu\nu}_r u_\nu\,,
	\qquad 
	\mu_h \coloneqq u^\mu \mc{A}_{r\mu}\,,
    \label{eq:sec3-def-E-muh}
\end{align}
and this satisfies $\mc{E}^\mu_r = h_r^{-1} E^\mu_r h_r$ as mentioned in the previous subsection.
$\mc{E}^\mu_r$ is the $r$-type gauge invariant matrix dressed in the color frame field $h_r$ and orthogonal to the velocity vector $\mc{E}^\mu_r u_\mu= 0$.
The (dressed) chemical potential contains the flow-direction derivative of the frame field $h_r^{-1} u \cdot \del h_r$ as 
\begin{align}
	\mu_h = u^\mu \left( h_r^{-1} A_{r\mu} h_r - \frac{\iu}{\gc} h_r^{-1} \del_\mu h_r \right)\,,
\end{align}
which gives the flow-direction response of the retained color frame.

The $h_r$ current contains the charge storage and the dissipative transverse response.
Around a color neutral state, the charge storage part becomes $\chi_h \mu_h u^\mu$, where $\chi_h$ is the color susceptibility of the environmental sector.\footnote{
In general, $\chi_h$ becomes an adjoint-space susceptibility matrix.
In this paper, we consider an anisotropic color-neutral background and the susceptibility is assumed to be proportional the identity matrix.
Its value is matching data, while local thermodynamic stability requires the corresponding static susceptibility matrix to be positive.
}
The dissipative transverse response is driven by the gauge-covariant thermodynamic force 
\begin{align}
	\mf{F}_h^\mu \coloneqq \mc{E}^\mu_r - T \Delta^{\mu\nu}_u \mc{D}_\nu[\mc{A}_r] \left( \frac{\mu_h}{T} \right)\,,
	\qquad 
	u_\mu \mf{F}_h^\mu = 0\,,
	\label{eq:sec3-h-fluid-force}
\end{align}
which is the standard covariant electrochemical force compatible with local thermality and stable diffusion.
Then, the local environmental $h_r$ current can be written at the first order of the derivative expansion as 
\begin{align}
	\mc{J}^\mu_{h,\fluid} = \chi_h \mu_h u^\mu + \sigma_h \mf{F}^\mu_h + \cdots 
	=\chi_h \mu_h u^\mu + \sigma_h \mc{E}^\mu_r - \sigma_h T \Delta^{\mu\nu}_u \mc{D}_\nu[\mc{A}_r] \left( \frac{\mu_h}{T} \right) + \cdots\,,
	\label{eq:sec3-h-fluid-current}
\end{align}
As seen in last of the previous subsection, the Ohmic contribution in the environmental current provides the friction term for the gauge field.\footnote{
In an isotropic color-neutral sate, the color conductivity is proportional to the identity matrix in the adjoint space.
The nondissipative antisymmetric responses, if present, are separate transport data.
In an isothermal constant-coefficient limit, we can write $D_{\mr{diff}}=\sigma_h/\chi_h$, but such limit is not assumed in this paper.}
Considering this $h_r$ current and the system current $J_\sys$, the EOM of the gauge field given by the variation of $A_{a\mu}$ becomes in the color frame 
\begin{align}
	h_r^{-1} D_\nu [A_r] F^{\nu\mu} h_r = \mc{J}^\mu_\sys + \mc{J}^\mu_{h,\fluid} =\mc{J}^\mu_\sys + \chi_h \mu_h u^\mu + \sigma_h \mf{F}^\mu_h + \cdots \,.
	\label{eq:sec3-h-frame-YM-dissipative}
\end{align}
This equation shows the local color response of the retained color frame environmental variables.
The charge-storage term $\chi_h\mu_hu^\mu$ is essential because it provides the nontrivial dissipation with the charge accumulation and the electrochemical relaxation between the system and the environment.
These currents enjoy the Ward identity
\begin{align}
	\mc{D}_\mu[\mc{A}_r] \left( \mc{J}^\mu_\sys + \mc{J}^\mu_{h,\fluid} \right) =0\,,
\end{align}
as \eqref{eq:sec2-total-Ward-id-stress-tensor} in the dissipative gravitational system.

\paragraph{Comments on noise and environmental SK action.}

The local noise associated with this dissipative response is fixed up to matching data by the SK/KMS constraints.
In the classical local Markovian limit, the dynamical KMS relates the local noise kernel to the dissipative part of the conductivity. The detailed derivation is shown in App.~\ref{app:dKMS-color-noise}. Indeed, for an isotropic color-neutral environment, the noise matrix is expressed as 
\begin{align}
	N^{\mu\nu}_{h,\mr{loc}} = 2 T \sigma_h \Delta^{\mu\nu}_u + \cdots\,,
	\label{eq:sec3-h-fluid-noise-FDT}
\end{align}
and the noise quadratic form is to be positive semi-definite because of the SK positivity.
These imply that $\sigma_h \geq 0$ in the isotropic environment.
In this sense, Eq.~\eqref{eq:sec3-h-fluid-force} is not merely a standard choice, but can also be regarded as natural from the EFT point of view, because it is organized by the symmetry associated with thermality.

To the order needed for the local dissipation and the noise, the environmental SK action can be written schematically as\footnote{
In the strict sense, the dressed gauge field $\mc{A}_{a\mu}$ is needed to be dressed in the gravitational $a$-type St\"uckelberg field as
\begin{align}
    \mbb{A}_{a\mu} \coloneqq \mc{A}_{a\mu} - \pounds_{X_a} \mc{A}_{r\mu}
\end{align}
as Eqs.~\eqref{eq:sec2-grav-Stuckelberg-replacement-g} and \eqref{eq:sec2-grav-Stuckelberg-replacement-phi} to keep the relative diffeomorphism invariance.
Then, the relative dressed gauge field in \eqref{eq:sec3-h-fluid-action} is replaced with $\mbb{A}_{a\mu}$ and the environmental SK action should be given by 
\begin{align}
	I_{h,\fluid} &= \frac{1}{2} \int \dd^dx \,T^{\mu\nu}_\env(u, T,\ldots) \mc{G}_{a\mu\nu} - 2 \int \dd^dx \Tr \mbb{A}_{a\mu} \mc{J}^\mu_{h,\fluid} (\mc{A}_r, \mu_h, u, T)
	\nn \\
	& \qquad 
	+ \iu \int \dd^dx \Tr \mbb{A}_{a\mu} (x) N^{\mu\nu}_{h,\mr{loc}}(x;\mc{A}_r, u, T) \mbb{A}_{a\nu}(x) 
	+ I_{\mr{hydro,KMS}}[u, T, \mc{G}_a] + \cdots\,.
	\label{eq:sec3-h-fluid-action-t}
\end{align}
}
\begin{align}
	I_{h,\fluid} &= \frac{1}{2} \int \dd^dx \,T^{\mu\nu}_\env(u, T,\ldots) \mc{G}_{a\mu\nu} - 2 \int \dd^dx \Tr \mc{A}_{a\mu} \mc{J}^\mu_{h,\fluid} (\mc{A}_r, \mu_h, u, T)
	\nn \\
	& \qquad 
	+ \iu \int \dd^dx \Tr \mc{A}_{a\mu} (x) N^{\mu\nu}_{h,\mr{loc}}(x;\mc{A}_r, u, T) \mc{A}_{a\nu}(x) 
	+ I_{\mr{hydro,KMS}}[u, T, \mc{G}_a] + \cdots\,.
	\label{eq:sec3-h-fluid-action}
\end{align}
The first term closes the energy-momentum exchange through the environment stress tensor and the second term closes the color excahange through the $h_r$ current.
The quadratic term of $\mc{A}_{a}$ is the local Markovian noise kernel.
In a gauge-covariant Hubbard--Stratnovich representation, the deterministic and stochastic currents obey the same total color Ward identity.
We therefore do not introduce a fundamental nonlocal noise kernel by dressing the fields on the two separated points with $h_r^{(-1)}$.
The last term $I_{\mr{hydro,KMS}}$ denotes the remaining hydrodynamic SK terms needed to impose the (dynamical) KMS condition and the positivity in the energy-momentum sector.

\paragraph{Integration of environmental variables.}

We consider integrating out the environmental variables by solving the Ward identity and getting the retarded solution.
The $a$-type St\"uckelberg variation leads to the Ward identity
\begin{align}
	\mc{D}_\mu[\mc{A}_r] \left( \mc{J}^\mu_\sys + \mc{J}^\mu_{h,\fluid} \right) =0\,.
	\label{eq:sec3-total-Ward-id}
\end{align}
Substituting the leading local current \eqref{eq:sec3-h-fluid-current} give the following color frame equation:
\begin{align}
	\mc{D}_\mu[\mc{A}_r] \left[ \mc{J}^\mu_\sys + \chi_h \mu_h u^\mu + \sigma_h \mf{F}^\mu_h \right] + \cdots = 0\,.
	\label{eq:sec3-h-fluid-local-Ward-expanded}
\end{align}
It is noted that we do not put any assumption that $\chi_h$, $\sigma_h$, $T$, $u^\mu$, or the hydrodynamic data are spacetime constants.
It is useful to define the local color frame operator by 
\begin{align}
	\hat{\msc{L}}_h \mu_h \coloneqq \mc{D}_\mu[\mc{A}_r] \left[
		\chi_h \mu_h u^\mu - \sigma_h T \Delta^{\mu\nu}_u \mc{D}_\nu[\mc{A}_r] \left( \frac{\mu_h}{T} \right)
	\right]\,,
	\label{eq:sec3-h-fluid-Lh-def}
\end{align}
and this is rewritten as 
\begin{align}
	\hat{\msc{L}}_h \mu_h = \mc{D}_\mu[\mc{A}_r] \left[ \mc{J}^\mu_{h,\fluid} - \sigma_h \mc{E}^\mu_r \right] + \cdots
	= - \mc{D}_\mu[\mc{A}_r] \left[ \mc{J}^\mu_\sys + \sigma_h \mc{E}^\mu_r \right] + \cdots\,\,
	\label{eq:sec3-h-fluid-local-eom-Lh}
\end{align}
which is the equation for the retained color frame response.
With the retarded boundary conditions, the formal solution is 
\begin{align}
	\bar{\mu}_h = - [\hat{\msc{L}}_h]^{-1}_{\mr{ret}} \mc{D}_\mu[\mc{A}_r] \left[
		\mc{J}^\mu_\sys + \sigma_h \mc{E}^\mu_r
	\right] + \cdots\,.
	\label{eq:sec3-h-fluid-muh-retarded-solution}
\end{align}
In the system-environment EFT description, the enlarged theory is local in the retained fields $(A_r, h_r,\ldots)$ and $\mu_h = u\cdot \mc{A}_r$ give the local response for the flow-direction dynamics of the color frame.
We note that we will have nonlocality after integrating the environmental field or equivalently substituting the solution of the EOM.

Substituting the retarded solution $\bar{\mu}_h$ in \eqref{eq:sec3-h-fluid-muh-retarded-solution} into the local frame current \eqref{eq:sec3-h-fluid-current} gives 
\begin{align}
	\mc{J}^\mu_{h} [\mc{A}_r] = \mc{J}^\mu_{h,\fluid}[\mc{A}_r, \bar{\mu}_h],
\end{align}
which is a retarded functional of the remaining soft fields and contains the inverse of the operator $\hat{\msc{L}}_h$.
In the original YM frame,\footnote{Following the hydrodynamics context, we often call this original YM frame as the physical color frame or the physical frame.} the corresponding current $J^\mu_{h,\fluid} = h_{r,\mr{ret}} \mc{J}^\mu_h[\mc{A}_r] h_{r,\mr{ret}}^{-1}$ is also nonlocal after integrating the color frame variables.
Thus the enlarged color frame sector is local and Markovian before integrating out the dressed chemical potential while the coarse-grained YM equation contains memory.
A further short-memory expansion requires an additional approximation.
If the retarded inverse is generated by a covariant transport operator, its physical-frame kernel can be represented by parallel transporters, which structure will appear explicitly in the hard-loop example in Sec.~\ref{sec:HTL-matching}.

In the physical color frame, the deterministic environmental current is 
\begin{align}
	J^\mu_{h,\fluid} = h_r \mc{J}^\mu_{h,\fluid} h_r^{-1}
	= h_r \left( \chi_h \mu_h u^\mu + \sigma_h \mf{F}^\mu_h + \cdots \right) h_r^{-1}\,.
	\label{eq:sec3-h-fluid-physical-current}
\end{align}
When the integration of the retained color response provides the noise term of $\mc{A}_a^2$, the reduced YM equation takes the stochastic Langevin form 
\begin{align}
	D_\nu[A_r] F^{\nu\mu}_r = J^\mu_\sys + J^\mu_{h,\fluid} + \eta^\mu_J\,.
	\label{eq:sec3-h-fluid-YM}
\end{align}
$\eta^\mu_J$ is the physical frame stochastic current obtained from the Hubbard-Stratonovich representation of the local color frame noise kernel.
This stochastic source is not an arbitrary external current, and it is constrained together with the other currents.
The total color Ward identity in the physical frame becomes 
\begin{align}
	D_\mu[A_r] \left( J^\mu_\sys + J^\mu_{h,\fluid} + \eta^\mu_J \right) = 0\,.
	\label{eq:sec3-h-fluid-total-Ward}
\end{align}
The Ward identity therefore constrains the full deterministic-plus-stochastic color current after the local environmental responses are included.

\paragraph{Ward identity for internal color symmetry and spacetime symmetry.}

Before closing this subsection, we give some comments on the Ward identity.
The color Ward identity discussed here should be distinguished from the spacetime Ward identity of the enlarged system.
The internal color Ward identity is enforced by the $a$-type color-frame field $\pi_a$ and constrains the total color currents.
If the spacetime response sector is also kept, the energy and momentum are carried by the hydrodynamic part of the environment.
When the YM field is included as a dynamical part of the closed enlarged system, we have schematically
\begin{align}
	\nabla_\mu \left( T^{\mu\nu}_\YM + T^{\mu\nu}_\sys + T^{\mu\nu}_\env \right) =0 \,,
	\label{eq:sec3-h-fluid-total-stress-Ward}
\end{align}
from the variation of the $a$-type St\"uckelberg field associated with the relative diffeomorphism.
This stress-tensor Ward identity is separate from Eq.~\eqref{eq:sec3-h-fluid-total-Ward}.  
The color exchange is closed by the dynamical color-frame sector, while the energy-momentum exchange is closed by the hydrodynamic stress tensor of the environment.

\subsection{Nonlocality from retained response variables}
\label{sec:color-frame-env-elimination}

An enlarged theory gives the local description of system-environment dynamics, which contains environmental variable so that introduce the dissipation and charge exchange between the system and environment.
The nonlocal retarded kernels appear only after these variables are solved with retarded boundary conditions and substituted back.
In this subsection, we formulate this general mechanism and relate it to the color-frame construction of the previous subsection.

In the color fluid example in Sec.~\ref{sec:color-fluid-completion}, the retained response is the flow-direction color frame variable $\mu_h$.
The environmental current can be expressed as 
\begin{align}
	\mc{J}^\mu_{h,\fluid} = \mc{J}^\mu_{h,\loc} + M^\mu_h \mu_h + \xi^\mu_h+  \cdots\,,
	\label{eq:sec3-Renv-color-current-split}
\end{align}
where we write
\begin{align}
	\mc{J}^\mu_{h,\loc} \coloneqq \sigma_h \mc{E}^\mu_r\,,
	\qquad
	M^\mu_h \mu_h \coloneqq \chi_h \mu_h u^\mu - \sigma_h T \Delta^{\mu\nu}_u \mc{D}_\nu[\mc{A}_r] \left( \frac{\mu_h}{T} \right) + \cdots\,.
\end{align}
$\xi^\mu_h$ is the local stochastic current associated with the quadratic SK noise kernel in Sec.~\ref{sec:color-fluid-completion}.
The covariant divergence of the total deterministic plus noise current gives the local response equation 
\begin{align}
	\hat{\msc{L}}_h \mu_h = S_h + \zeta_h\,,
	\label{eq:sec3-Renv-color-response-eq}
\end{align}
where we introduce
\begin{align}
	S_h \coloneqq - \mc{D}_\mu [\mc{A}_r] \left( \mc{J}^\mu_\sys + \sigma_h \mc{E}^\mu_r \right)\,,
	\qquad 
	\zeta_h \coloneqq - \mc{D}_\mu[\mc{A}_r] \xi^\mu_h\,.
	\label{eq:sec3-Renv-zetah-def}
\end{align}
Thus, $\zeta_h$ in \eqref{eq:sec3-Renv-color-response-eq} is not an additional nonlocal noise but the source induced by the local current noise and the Ward identity.

With the retarded boundary conditions, the formal solution of \eqref{eq:sec3-Renv-color-response-eq} is given by 
\begin{align}
	\bar{\mu}_h = [\hat{\msc{L}}^{-1}_h]_{\mr{ret}} (S_h + \zeta_h)\,,
	\label{eq:sec3-Renv-muh-retarded}
\end{align}
and we obtain the following equation by substituting this solution back into the environmental current 
\begin{align}
	\mc{J}^\mu_{h,\fluid} = \mc{J}^\mu_{h,\loc} + M^\mu_h [\hat{\msc{L}}_h^{-1}]_{\mr{ret}} S_h + \Bigl[ \xi^\mu_h + M^\mu_h [\hat{\msc{L}}_h^{-1}]_{\mr{ret}} \zeta_h \Bigr] + \cdots\,.
	\label{eq:sec3-Renv-color-current-after-elimination}
\end{align}
The second term is the induced retarded memory current.
The third bracketed terms are the reduced stochastic current.
Both nonlocal contributions arise from a local enlarged description.

The same structure can be considered without referring to the specific color fluid channel.
Let $\mc{R}^I_\env$ denote retained environmental responses with $I$ including all discrete and continuous labels.
A local Markovian response equation has the following form
\begin{align}
	\hat{\msc{L}}_{\env, IJ} \mc{R}^J_\env &= S_{\env,I}[\mc{A}_r] + \zeta_{\env,I}\,,
	\label{eq:sec3-Renv-eq}
	\\
	\mc{J}^\mu_{\env}&= \mc{J}^\mu_\loc[\mc{A}_r] + M^\mu_I[\mc{A}_r] \mc{R}^I_\env + \xi^\mu_\env\,.
	\label{eq:sec3-Renv-local-template}
\end{align}
These are generalized equations of \eqref{eq:sec3-Renv-color-current-split} and \eqref{eq:sec3-Renv-color-response-eq}.
We write the inverse of the operator with the retarded boundary conditions as 
\begin{align}
	G^{IJ}_R \coloneqq [\hat{\msc{L}}_\env^{-1}]^{IJ}_{\mr{ret}}\,,
\end{align}
and the formal solution of \eqref{eq:sec3-Renv-eq} is 
\begin{align}
	\bar{\mc{R}}_\env^I = G^{IJ}_R \Bigl(
		S_{\env,J}[\mc{A}_r] + \zeta_{\env, J}
	\Bigr)\,.
\end{align}
Substituting this solution into \eqref{eq:sec3-Renv-local-template}, we obtain
\begin{align}
	\mc{J}^\mu_\env = \mc{J}^\mu_\loc[\mc{A}_r] + M^\mu_I[\mc{A}_r] G^{IJ}_R S_{\env,J}[\mc{A}_r] + \Bigl[
		\xi^\mu_\env + M^\mu_I G^{IJ}_R \zeta_{\env,J}
	\Bigr]\,.
	\label{eq:sec3-Renv-current-after-elimination}
\end{align}
This is also the general form of \eqref{eq:sec3-Renv-color-current-after-elimination}.
The second term is deterministic and retarded, while the third bracketed terms are contributions from the reduced noise.

For the color fluid, the dictionary is
\begin{align}
	\mc{R}_{\env}
	&\leftrightarrow
	\mu_h\,,
	&
	\hat{\msc{L}}_{\env}
	&\leftrightarrow
	\hat{\msc{L}}_h\,,
	&
	S_{\env}
	&\leftrightarrow
	S_h\,,
	&
	\zeta_{\env}
	&\leftrightarrow
	\zeta_h\,,
	\nn\\
	\mc{J}_{\loc}^\mu
	&\leftrightarrow
	\sigma_h \mc{E}_r^\mu\,,
	&
	M^\mu \mc{R}_{\env}
	&\leftrightarrow
	M_h^\mu\mu_h \,.
	\label{eq:sec3-Renv-color-dictionary}
\end{align}
In the hard-loop application, the analogous retained response is the velocity-resolved hard-sector variable $W_r(x,v)$ with 
\begin{align}
	\mc{R}^I_\env \to W_r(x,v)\,,
	\quad 
	\hat{\msc{L}}_\env \to v \cdot \mc{D}[\mc{A}_r] + C\,,
	\quad 
	S_\env \to v \cdot \mc{E}_r
	\quad
\text{up to matching normalization}\,.
	\label{eq:sec3-Renv-hard-loop-dictionary}
\end{align}

Finally, we note that this elimination of the environmental variables should be distinguished from a further soft-only projection of the color-frame sector itself.
Eliminating $\mu_h$ or $W_r(x,v)$ produces a retarded induced current while the color-frame completion $(h_r,\pi_a)$ can still be kept in the enlarged local description. 
A further reduction can also solve or fix the remaining color frame. 
We write the resulting representative schematically by
\begin{align}
	h_{r,\mr{ret}}(x) = h_{r,\mr{ret}}[A_r, \mr{matter},\ldots](x)\,.
	\label{eq:sec3-h-ret-def}
\end{align}
The symbol $h_{r,\mr{ret}}$ is not an additional dynamical field in the reduced theory.
It records how the environmental color frame is related to the physical color frame after the final soft-only projection. 
The gauge covariance requires this representative to transform equivalently as 
\begin{align}
	h_{r,\mr{ret}}[A_r^{U_r},\mr{matter}^{U_r},\ldots] (x) = U_r(x) h_{r,\mr{ret}}[A_r, \mr{matter},\ldots] (x)\,.
\end{align}
With this convention, a bilocal kernel first written in the environmental color frame is converted back to the physical color frame as 
\begin{align}
	K^{\mr{phys}}_R (x,y;A_r) = h_{r,\mr{ret}}(x) K^{\env}_R(x,y; \mc{A}_r[h_{r,\mr{ret}}]) h_{r,\mr{ret}}^{-1}(y)\,.
	\label{eq:sec3-physical-bilocal-kernel}
\end{align}
The endpoint factors are therefore color-basis rotations associated with the final projection of the color frame. 
They are not additional nonlocal interactions in the original enlarged local action. 
The explicit Wilson lines appear only when the eliminated response is governed by a covariant transport operator, as in the hard-loop example in the next section. 
Thus the two operations are distinct: $\mu_h$ or $W_r(x,v)$ is eliminated to generate a retarded induced response, while $h_r$ is solved or fixed only in a further soft-only reduction.

\subsection{Dressed fermions and induced environmental damping}
\label{sec:gauge-invariant-fermions}

The fermionic matter sector is an illustrative example of the matter sector in the system motivated by QCD.
The purpose of this section is to show that the same color frame approach is also applied to the matter sectors so that we locally describe the interaction between the system and the environment.
Integrating out the retained response then produces a gauge-covariant retarded self-energy, whose short-memory limit gives the local damping.

Let us consider a Dirac fermion $\psi$ being a fundamental representation of the gauge group $G$.
Under the usual gauge transformation by $U \in G$, these behave as 
\begin{align}
	\psi \mapsto U \psi\,,
	\qquad 
	\bar{\psi} \mapsto \bar{\psi} U^{-1}\,.
	\label{eq:sec3-fermion-transform}
\end{align}
The classical action is given by
\begin{align}
	S_\psi = - \int \dd^d x\, \bar{\psi} \left[ \gamma^\mu D_\mu[A] + m \right] \psi
	= - \int \dd^d x\, \bar{\psi} \left[ \sld{D}[A] + m \right] \psi\,,
\end{align}
where the covariant derivative acts on $\psi$ as $D_\mu[A] \psi = \del_\mu \psi + \iu \gc A_\mu \psi$.
We write the action of $\sld{D}[A]$ on $\bar{\psi}$ as 
\begin{align}
	\bar{\psi} \overleftarrow{\sld{D}}[A] = (\del_\mu \bar{\psi} - \iu \gc \bar{\psi} A_\mu) \gamma^\mu \,.
\end{align}
Following the same procedure with the gauge field, we define the dressed fermion on the SK branches by 
\begin{align}
	\varPsi_\pm \coloneqq h^{-1}_\pm \psi\,,
	\qquad 
	\bar{\varPsi}_\pm \coloneqq \bar{\psi} h_\pm\,,
	\qquad 
	h_\pm = h_r \exp \left(\pm \frac{\iu}{2} \pi_a \right)\,,
	\label{eq:sec3-dressed-fermion-pm}
\end{align}
and the fermions in the $r$-$a$ basis are introduced by 
\begin{align}
	\varPsi_r \coloneqq \frac{1}{2}(\varPsi_+ + \varPsi_-)\,,&
	\qquad 
	\varPsi_a \coloneqq \varPsi_+ - \varPsi_-\,,
	\\
	\bar\varPsi_r \coloneqq \frac{1}{2}(\bar\varPsi_+ + \bar\varPsi_-)\,,&
	\qquad 
	\bar\varPsi_a \coloneqq \bar\varPsi_+ - \bar\varPsi_-\,.
\end{align}
At the leading order in the $a$-type fields, the closed fermion SK action becomes
\begin{align}
	I_{\psi,0} = - \int \dd^dx \, \left[
        \bar{\varPsi}_a \left(\sld{\mc{D}}[\mc{A}_r] + m \right) \varPsi_r + \bar{\varPsi}_r  \left(
            - \overleftarrow{\sld{\mc{D}}}[\mc{A}_r] +m 
        \right) \varPsi_a
    \right] - 2 \int \dd^dx \Tr\mc{A}_{a\mu} \mc{J}^{\mu}_{\varPsi},
\end{align}
where the current is given by 
\begin{align}
    \mc{J}^\mu_\varPsi = \mc{J}^{A\mu}_\varPsi T^A\,,
    \qquad 
    \mc{J}^{A\mu}_\varPsi = \iu \gc \bar{\varPsi}_r \gamma^\mu T^A \varPsi_r\,.
\end{align}
The corresponding current in the original YM frame is 
\begin{align}
	J_\varPsi^\mu = h_r \mc{J}^\mu_\varPsi h_r^{-1}\,.
\end{align}
The dressed covariant derivative acts as 
\begin{align}
	\sld{\mc{D}}[\mc{A}_r] \varPsi_r = \gamma^\mu (\del_\mu + \iu \gc \mc{A}_{r\mu}) \varPsi_r\,,
	\qquad 
	\bar{\varPsi}_r \overleftarrow{\sld{\mc{D}}}[\mc{A}_r] = (\del_\mu \bar{\varPsi} - \iu \gc \bar{\varPsi} \mc{A}_\mu) \gamma^\mu\,,
\end{align}
and satisfies 
\begin{align}
	h_r^{-1} \sld{D}[A_r] \psi_r = \sld{\mc{D}}[\mc{A}_r] \varPsi_r\,,
	\qquad 
	\bar{\psi}_r \overleftarrow{\sld{D}}[A_r] h_r = \bar{\varPsi}_r \overleftarrow{\sld{\mc{D}}}[\mc{A}_r]\,.
\end{align}

Before integrating out the environmental variables, the interactions with the color frame environment are local in the retained variables as seen in this section.
The leading bilinear response in the color frame can be written as the additive self-energy in the inverse of the fermion propagator
\begin{align}
	I_{h\psi} = - \int \dd^dx  \left(
		\bar{\varPsi}_a \mc{M}_\psi[\mc{A}_r, h_r] \varPsi_r + \bar{\varPsi}_r \mc{M}_{\psi,A}[\mc{A}_r, h_r] \varPsi_a \right) + I^\loc_{\psi,K}\,,
	\label{eq:sec3-fermion-local-h-coupling}
\end{align}
where $I^\loc_{\psi,K}$ is the local Keldysh term.
The leading operator in $\mc{M}_\psi$ is 
\begin{align}
	\mc{M}_\psi[\mc{A}_r, h_r] = -\gamma_\psi(\mu_h, T,\ldots) \sld{u} + \mc{O}(\mc{D}[\mc{A}_r], \mc{F}_r)\,, 
	\label{eq:sec3-fermion-local-Mpsi}
\end{align}
where $\mc{O}(\mc{D}[\mc{A}_r], \mc{F}_r)$ denotes the higher-derivative and field strength corrections.
Since the chemical potential $\mu_h$ contains $h_r^{-1} u\cdot \del h_r$, this operator gives a local coupling to the dynamical color frame sector.
In a color-neutral isotropic background, we can expand the coefficient as 
\begin{align}
	\gamma_\psi(\mu_h, T,\ldots) = \gamma_{\psi,0}(T) + \gamma_{\psi,2}(T) \Tr \mu_h^2+ \cdots\,.
\end{align}
The advanced part in \eqref{eq:sec3-fermion-local-h-coupling} is fixed by SK reality 
\begin{align}
	\mc{M}_{\psi,A} = - \gamma^0 \mc{M}_\psi^\dagger \gamma^0\,.
\end{align}
With the convention in \eqref{eq:sec3-fermion-local-h-coupling}, the retarded and advanced local self-energies are the kernels added to the inverse of the fermion propagators 
\begin{align}
	\Sigma^\loc_R \coloneqq \mc{M}_\psi[\mc{A}_r,h_r]\,,
	\qquad 
	\Sigma^\loc_A \coloneqq \mc{M}_{\psi,A}[\mc{A}_r, h_r]\,,
\end{align}
which are related by SK reality as
\begin{align}
	\Sigma^\loc_A = - \gamma^0 \left( \Sigma^\loc_R \right)^\dagger \gamma^0\,.
\end{align}
Thus, the leading Markovian damping term is given by 
\begin{align}
	\Sigma^\loc_R \approx - \gamma_\psi \sld{u}\,,
	\qquad 
	\Sigma_A^\loc \approx \gamma_\psi \sld{u}\,.
\end{align}

The local Keldysh term $I^\loc_{\psi,K}$ is the fermionic counterpart of the noise kernel and can be written schematically as 
\begin{align}
	I_{\psi,K}^\loc = \int \dd^d x \bar{\varPsi}_a \Sigma^\loc_K \varPsi_a\,.
\end{align}
$\Sigma^\loc_K$ is the local Keldysh self-energy kernel induced by the environment.
In thermal equilibrium, $\Sigma^\loc_K$ is not an independent coefficient relates the self-energies.
In frequency space along the bath frame, with $\omega$ conjugate to the proper time in the local rest frame of $u^\mu$, the fermionic fluctuation-dissipation relation is 
\begin{align}
	\Sigma_K(\omega) = \left[ 1 - 2 n_F(\omega) \right] (\Sigma_R(\omega) - \Sigma_A(\omega))\,,
\end{align}
where 
\begin{align}
	n_F(\omega) = \frac{1}{\ee^{\beta \omega} + 1 }\,,
\end{align}
is the Fermi--Dirac distribution in the bath frame.
The local kernel $\Sigma^\loc_K$ should be understood as the leading contribution in the derivative expansion of this equilibrium Keldysh self-energy.
The passivity requires the anti-hermitian part of $\Sigma_R$ to damp positive-norm excitations, which gives $\gamma_\psi \geq 0$ in the local rest frame.

\paragraph{Integration of environmental variables.}

After the retained response in the environmental sector is solved with the retarded boundary conditions and substituted back, the fermion obtains the induced nonlocal self-energy kernels.
In the original physical frame, the induced contribution has the following schematic form 
\begin{align}
	I_{\psi,\ind} = - \int \dd^dx \dd^d y \left[
		\bar{\psi}_a(x) \Sigma^\ind_R(x,y;A_r) \psi_r(y) + \bar{\psi}_r(x) \Sigma^\ind_A(x,y;A_r) \psi_a(y)
	\right] + I^\ind_{\psi,K}\,,
	\label{eq:sec3-fermion-induced-nonlocal}
\end{align}
where $\Sigma^\ind_{R,A}(x,y;A_r)$ are the induced retarded and advanced fermion self-energy kernels, and $I^\ind_{\psi,K}$ is the corresponding induced Keldysh part.
The retarded kernel contains the retarded Green function of the integrated response in the environmental sector, so the nonlocality appears only after the retained environmental response is integrated out.

In the short-memory regime, the induced kernel reduces to the same local structure as 
\begin{align}
	\Sigma^\ind_R(x,y) \to \mc{M}_\psi \delta^d(x-y) + \text{derivative corrections}\,.
\end{align}
This is a further approximation made after the integration of the retained variables.
Using the leading term in \eqref{eq:sec3-fermion-local-Mpsi}, the local short-memory equation in the environmental color frame becomes 
\begin{align}
	(\sld{\mc{D}}[\mc{A}_r] +m - \gamma_\psi \sld{u}) \varPsi_r =0\,.
	\label{eq:sec3-fermion-friction-eom}
\end{align}
The corresponding Keldysh self-energy controls fermionic correlations, which provides a classical Langevin force but is not written here.
In the rest frame, \eqref{eq:sec3-fermion-friction-eom} contains $\gamma^0 (\del_t + \gamma_\psi)$ and $\gamma_\psi>0$ gives the damping.

The damping term produces the color exchange with the integrated response in the environmental sector, which is not an isolated nonconserved color sink.
In the local damping approximation, the fermionic current need not be conserved by itself, but its nonconserved contribution should be compensated by the color frame environment as 
\begin{align}
	\mc{D}_\mu[\mc{A}_r] \left( \mc{J}^\mu_\varPsi + \mc{J}^\mu_{h,\psi} \right) = 0\,.
\end{align}
Here, $\mc{J}^\mu_{h,\psi}$ denotes the compensating environmental color current induced.
In the physical frame, this Ward idendity becomes 
\begin{align}
	D_\mu[A_r]\left( J^\mu_\varPsi + J^\mu_{h,\psi} \right) = 0\,,
\end{align}
where $J^\mu_{h,\psi} = h_r \mc{J}^\mu_{h,\psi} h_r^{-1}$.
Thus fermion damping is the system part of a color exchange with the same dynamical color frame environment.

\section{Hard-loop response from local hard sector EFT}
\label{sec:HTL-matching}

In this section, we apply the local system-environment EFT constructed in Sec.~\ref{sec:NA-gauge} to a concrete benchmark: the hard-thermal-loop (HTL) response of a hot non-Abelian plasma. 
In this example, the soft YM field interacts with hard colored excitations. 
After these hard excitations are integrated out, their collective color response appears in the soft theory as a nonlocal hard-loop induced by the current. 
Our purpose is to derive this nonlocal structure from a local EFT.

In order to derive the nonlocality, we keep the $h_r$ sector introduced in Sec.~\ref{sec:NA-gauge} and add a velocity-resolved hard response variable $W_r(x,v)$ with the hard velocity $v^\mu$.
The $h_r$ sector supplies the dressed connection $\mathcal A_r[h_r]$, which allows the hard response to be described in local gauge-covariant variables. 
The field $W_r(x,v)$ describes how hard colored excitations moving with velocity $v^\mu$ respond to the soft gauge field. The local EFT is therefore written in terms of $\mathcal A_r[h_r]$, the retained response $W_r(x,v)$, and the corresponding local $a$-type sources. 
After the local $W_r$ equation is solved with the retarded Green function, the induced hard current becomes a nonlocal functional of the soft gauge field. In the collisionless limit, this current reproduces the standard HTL response. Thus the hard-loop nonlocality is not inserted directly into the soft YM theory; it arises from the retarded propagation of the retained hard response.

In Sec.~\ref{subsec:sec4-velocity-markov}, we define the local velocity-space response $W_r(x,v)$, its local SK action, and the hard-current Ward identity. 
In Sec.~\ref{subsec:sec4-retarded-projection}, we integrate out this response at fixed $h_r$ and obtain the retarded nonlocal current, together with its zero-mode-preserving short-memory limit. 
In Sec.~\ref{subsec:sec4-noise-kms}, we construct the corresponding fluctuation sector, including the local $W$-sector noise, positivity, and thermal matching. 
Finally, Sec.~\ref{subsec:sec4-HTL-benchmark} takes the collisionless limit and shows that the retarded current obtained by integrating out $W_r(x,v)$ reduces to the standard HTL induced current written in ordinary YM variables.

\subsection{Local velocity space hard response}
\label{subsec:sec4-velocity-markov}

This subsection defines the local hard-response sector before $W_r$ is integrated out. We introduce the velocity label, the retained response $W_r(x,v)$, and a local SK action whose $\pi_a$ equation imposes the hard-current Ward identity. 
The result is a local theory in the variables $(A_r,h_r,W_r)$, which will be matched at the end of this subsection to the general environmental response template of Sec.~\ref{sec:NA-gauge}.

For the HTL benchmark, we use the same environmental velocity field $u^\mu$ ($u^2 = -1$) as in Sec.~\ref{sec:NA-gauge} and assume that it specifies the local rest frame of a hot plasma.
A future-directed hard velocity is written with a normalized spatial vector $n^\mu$ as
\begin{align}
v^\mu \coloneqq u^\mu+n^\mu\,,
\qquad
u\cdot n=0\,,
\qquad
n^2=1 \,.
\label{eq:sec4-velocity-def}
\end{align}
For an arbitrary function $f(v)$, the normalized angular average over hard directions in this local rest frame is
\begin{align}
\int_v f(v)
\coloneqq
\int\frac{\dd\Omega_{d-2}}{\Omega_{d-2}}
f\!\left(u+n\right)\,,
\label{eq:sec4-velocity-integral-def}
\end{align}
where $\dd\Omega_{d-2}$ is the solid-angle element on the unit sphere $S^{d-2}$ in the spatial directions orthogonal to $u^\mu$, and $\Omega_{d-2}$ is its total solid angle. 
For this angular average, we find
\begin{align}
\int_v 1=1\,,
\qquad
\int_v n^\mu=0\,,
\qquad
\int_v n^\mu n^\nu=\frac{1}{d-1}\Delta_u^{\mu\nu}\,,
\qquad
\Delta_u^{\mu\nu}\coloneqq g^{\mu\nu}+u^\mu u^\nu \,.
\label{eq:sec4-velocity-average}
\end{align}
The velocity labels an internal kinetics, and the average $\int_v$ is local in spacetime.

The next step is to specify the local hard-sector variable that will be retained before taking the HTL limit. 
In a hot plasma, hard excitations are resolved not only by the spacetime point $x$ but also by their propagation direction $v^\mu$. 
A soft YM perturbation therefore changes the distribution of the excitations in a direction-dependent way. 
We encode this directional linear response by a matrix-valued local variable $W_r(x,v)$. 
The field $h_r$ is not an additional source for this response. 
The role of $h_r$ is to form the dressed connection $\mathcal A_r[h_r]$ so that the hard response is transported covariantly.
The perturbation enters through the color electric field built from this dressed connection in \eqref{eq:sec3-def-E-muh}.
For a velocity mode $v^\mu$, the corresponding local source is $v\cdot\mathcal E_r$.

We now write the local linear equation for the retained response. 
Since $W_r$ is resolved in velocity space, the kinetic operator acts both on spacetime and on the velocity label. 
Locality in spacetime forbids the operator from connecting different spacetime points, while local hard-sector interactions may still mix different velocity directions at the same point $x$. 
The part fixed by gauge-covariant propagation is given by the derivative along the hard trajectory, $v\cdot\mathcal D[\mathcal A_r]W_r(x,v)$.
The remaining local linear mixing in velocity space is parametrized by a collision kernel $C_{vv'}$. 
We therefore define the kinetic operator $\mc{K}$ action on $W_r$ by
\begin{align}
(\mathcal K W_r)(x,v)
\coloneqq
v\cdot \mathcal{D}[\mathcal{A}_r] W_r(x,v)
+
\int_{v'}C_{vv'}W_r(x,v') \,.
\label{eq:sec4-K-operator}
\end{align}
Here $C_{vv'}$ is local in spacetime but acts on the velocity label and adjoint color indices.
For a color-neutral environment, it is an invariant tensor in adjoint color space.
The deterministic local equation for the hard response is then given by
\begin{align}
(\mathcal K W_r)(x,v)
=
v\cdot\mathcal{E}_r(x)\, .
\label{eq:sec4-W-Markov-eom-det}
\end{align}

A collisional Markov completion may provide a local stochastic force $\zeta_W(x,v)$,
\begin{align}
(\mathcal K W_r)(x,v)
=
v\cdot\mathcal{E}_r(x)+\zeta_W(x,v) \,.
\label{eq:sec4-W-Markov-eom}
\end{align}
The collision kernel and the stochastic force are not arbitrary. 
Both preserve the velocity zero mode required by the covariant Ward identity as derived below. 
In the strictly collisionless HTL benchmark, one sets $C_{vv'}=0$ and omits the local bulk noise, so that the retarded response is generated solely by the gauge-covariant propagation term $v\cdot\mathcal D[\mathcal A_r]$. 
The hard susceptibility is denoted by $m_D^2$, which is the Debye mass parameter in the collisionless HTL benchmark.
Before integrating out $W_r$, the hard-sector current is the local velocity moment
\begin{align}
\mathcal J_{\rm hard}^\mu(x)
\coloneqq
m_D^2\int_v v^\mu W_r(x,v) \,.
\label{eq:sec4-hard-current-W}
\end{align}
This current becomes nonlocal only after the local equation for $W_r$ is solved using the retarded Green function of the operator $\mathcal K$.

A compact local SK action for this retained hard sector keeps three ingredients together. 
It contains the coupling of the hard current to the $a$-type source, the local response equation for $W_r$, and the local Keldysh kernel for the retained hard response. 
From the EFT viewpoint, this is the minimal leading order local structure compatible with the retained variables, the $a$-field expansion, gauge covariance, and the velocity space derivative counting. 
Here $W_a$ enforces the local response equation, $\mathcal A_{a\mu}$ couples to the induced hard current, and the most general local quadratic $a$-$a$ term is encoded in $N^{\rm loc}_{vv'}$. 
To quadratic order in the $a$-fields, the hard-sector contribution may be written schematically as
\begin{align}
I_{\rm hard}^{\rm loc}
=&
-2\int \dd^dx\,
\Tr\left[
\mathcal A_{a\mu}\mathcal J_{\rm hard}^\mu[W_r]
\right]
+
2\int \dd^dx\int_v
\Tr\left[
W_a(x,v)
\left(
(\mathcal K W_r)(x,v)-v\cdot\mathcal{E}_r(x)
\right)
\right]
\nonumber\\
&+
\iu\int \dd^dx\int_v\int_{v'}
\Tr\left[
W_a(x,v)N^{\rm loc}_{vv'}(x)W_a(x,v')
\right]
+\cdots\, .
\label{eq:sec4-local-W-action}
\end{align}
The omitted terms include higher order $a$-field terms and possible higher gradient local operators allowed by the same symmetries. 
With the choice above, its deterministic equation gives \eqref{eq:sec4-W-Markov-eom-det}. 
The last term is the local Keldysh term for the retained hard response. 
The kernel $N^{\rm loc}_{vv'}$ will be constrained by the Ward identity, SK positivity, and thermal matching, as discussed in Sec.~\ref{subsec:sec4-noise-kms}.

The coupling of $\mc{A}_{a\mu} \mc{J}^\mu_{\mr{hard}}$ imposes the Ward identity on the hard sector current.
Keeping the part of $\mathcal A_{a\mu}$ linear in the relative gauge variable $\pi_a$, the first term in \eqref{eq:sec4-local-W-action} contains
\begin{align}
I_{\rm hard}^{\rm loc}
\ni
\frac{2}{\gc}\int\dd^dx\,
\Tr\left[
\pi_a\, \mathcal{D}_\mu[\mathcal{A}_r] \mathcal{J}^\mu_{\rm hard}
\right]
+
\cdots\, .
\label{eq:sec4-pia-hard-ward-coupling}
\end{align}
Thus the variation of $\pi_a$ imposes the local hard-current Ward identity
\begin{align}
\mathcal{D}_\mu[\mathcal{A}_r] \mathcal{J}^\mu_{\rm hard}=0\, .
\label{eq:sec4-hard-current-Ward-target}
\end{align}
On the other hand, the covariant divergence of $\mc{J}^\mu_{\mr{hard}}$ is evaluated using \eqref{eq:sec4-W-Markov-eom} and \eqref{eq:sec4-hard-current-W} at leading gradient order as
\begin{align}
\mathcal{D}_\mu[\mathcal{A}_r] \mathcal{J}^\mu_{\rm hard}
=
m_D^2\int_v v\cdot\mathcal{E}_r
+
m_D^2\int_v\zeta_W(x,v)
-
m_D^2\int_v\int_{v'}C_{vv'}W_r(x,v')\, .
\label{eq:sec4-hard-Ward-C-intermediate}
\end{align}
The first term vanishes 
\begin{align}
    \int_v v \cdot \mc{E}_r = 0\,,
\end{align}
because $v^\mu=u^\mu+n^\mu$, $u\cdot\mathcal E_r=0$, and $\int_v n^\mu=0$. Compatibility with \eqref{eq:sec4-hard-current-Ward-target} for arbitrary $W_r$ then requires the collision kernel to annihilate the velocity zero mode. The stochastic Ward identity imposes the corresponding condition on the noise:
\begin{align}
\int_v C_{vv'}=0\,,
\qquad
\int_v\zeta_W(x,v)=0\, .
\label{eq:sec4-zero-mode-constraints}
\end{align}
These conditions do not forbid collisions. They restrict the present minimal hard-sector completion to charge-conserving collisions within the $W_r$ sector. 
Collisions may redistribute the hard response among velocity directions and relax non-uniform angular components, but they cannot relax the velocity average $\int_v W_r$. 
If collisions are allowed to transfer net color charge to another charge-carrying sector, we need to keep that sector explicitly so that the total current, rather than $\mathcal J_{\rm hard}^\mu$ alone, satisfies the covariant Ward identity.

With these ingredients, the retained hard sector is a specialization of the general environmental-response template of Sec.~\ref{sec:color-frame-env-elimination}. 
The correspondence is
\begin{align}
\mathcal R_{\rm env}^I
&\longrightarrow
W_r(x,v)\,,
&
I
&\longrightarrow
({\rm hard},v)\,,
\nonumber\\
\hat{\mathcal{L}}_{{\rm env},IJ}
&\longrightarrow
v\cdot\mathcal D[\mathcal A_r]\delta_{vv'}+C_{vv'}\,,
&
S_{{\rm env},I}
&\longrightarrow
v\cdot\mathcal{E}_r\,,
\nonumber\\
\mathcal J_{\rm loc}^\mu
&\longrightarrow
0\,,
&
M_I^\mu\mathcal R_{\rm env}^I
&\longrightarrow
m_D^2\int_v v^\mu W_r(x,v)\,.
\label{eq:sec4-hard-loop-specialization}
\end{align}
In the next subsection, we introduce the velocity zero-mode decomposition and integrate out $W_r(x,v)$ using the retarded Green function of $\mathcal K$ while keeping $h_r$ fixed.

\subsection{Retarded hard current and short-memory limit}
\label{subsec:sec4-retarded-projection}

This subsection applies the derivation of the nonlocality in Sec.~\ref{sec:NA-gauge} to the HTL hard response. 
A local system-environment EFT becomes nonlocal in the system variables only after a retained environmental response is integrated out with a retarded Green function. 
In the current situation,  the retained response is the velocity-resolved hard variable $W_r(x,v)$. 
We solve its local equation using the retarded Green function of $\mathcal K$ and substitute the result into the local current $\mathcal J_{\rm hard}^\mu=m_D^2\int_v v^\mu W_r$ as similar way in Sec.~\ref{sec:color-fluid-completion}.
This gives a retarded nonlocal hard current as a functional of the dressed connection $\mathcal A_r[h_r]$, while $h_r$ itself is still kept explicit.
We then take a short-memory limit in the part of the velocity space response that is not protected by the Ward identity.

The identity kernel in velocity space is denoted by $\delta_{vv'}$ and is defined with respect to the normalized measure $\int_v$ by
\begin{align}
\int_{v'}\delta_{vv'}f(v')=f(v)\,.
\end{align}
The retarded Green function of the local transport operator is
\begin{align}
G_R^W(x,v;y,v')
\coloneqq
\left[
v\cdot\mathcal D[\mathcal A_r]\delta_{vv'}+C_{vv'}
\right]^{-1}_{\rm ret}(x,v;y,v')\,,
\label{eq:sec4-GR-def}
\end{align}
and the deterministic solution is
\begin{align}
W_{r,{\rm det}}(x,v)
=
\int_{v'}\int\dd^dy\,
G_R^W(x,v;y,v')\,
v'\cdot\mathcal{E}_r(y)\, .
\label{eq:sec4-W-det-solution}
\end{align}
Substituting this solution into \eqref{eq:sec4-hard-current-W} gives
\begin{align}
\mathcal J_{{\rm hard},{\rm det}}^\mu(x)
=
m_D^2
\int_v\int_{v'}\int\dd^dy\,
v^\mu
G_R^W(x,v;y,v')\,
v'\cdot\mathcal{E}_r(y)\, .
\label{eq:sec4-hard-current-retarded}
\end{align}
This is the retarded nonlocal current generated by integrating out the retained hard response. In the original local EFT, the current is local in $W_r$. 
The nonlocal dependence on $\mathcal E_r(y)$ appears only after the $W_r$ equation is solved with the retarded Green function $G_R^W$. 
The stochastic part is obtained by the same retarded propagation with $v'\cdot\mathcal{E}_r$ replaced with $\zeta_W$.

We next separate the velocity average of $W_r$ from its direction-dependent part. 
Here, the zero mode means the component of the velocity space function $W_r(x,v)$ which is independent of the hard direction $v$. 
We note that it is not a zero spacetime momentum mode. 
The projector onto this velocity-space zero mode and its complement are
\begin{align}
(P_0f)(x,v)
\coloneqq
\int_{v'}f(x,v')\,,
\qquad
P_\perp\coloneqq 1-P_0 \,.
\label{eq:sec4-P0-Pperp}
\end{align}
Thus $P_0f$ is independent of $v$, while $P_\perp f$ contains only the direction-dependent angular components. 
In kernel notation, we write
\begin{align}
P_{\perp,vv'}=\delta_{vv'} - P_{0,vv'} \,,
\label{eq:sec4-Pperp-kernel}
\end{align}
where $P_{0,vv'}$ is the rank-one kernel that maps a function to its velocity average. 
We decompose the retained hard response as
\begin{align}
W_0(x)
\coloneqq
\int_v W_r(x,v)\,,
\qquad
W_\perp(x,v)
\coloneqq
P_\perp W_r(x,v) \,.
\label{eq:sec4-W0-Wperp}
\end{align}
The zero mode $W_0$ is the velocity-averaged part of the hard response, which has a constant profile in the velocity space.
Using $v^\mu=u^\mu+n^\mu$ and $\int_v n^\mu=0$, the hard current is decomposed as
\begin{align}
\mathcal J_{\rm hard}^\mu
=
m_D^2u^\mu W_0
+
m_D^2\int_v n^\mu W_\perp(v) \, .
\label{eq:sec4-current-zero-mode-decomp}
\end{align}
This is the hard sector analogue of the current decomposition in \eqref{eq:sec3-h-fluid-current}. 
The coefficient of $u^\mu$ is the charge-storage part, and the remaining term is the spatial current in the local plasma rest frame. 
Thus $m_D^2W_0$ plays the role of the hard-sector color density, in analogy with $\chi_h\mu_h$ in Sec.~\ref{sec:color-fluid-completion}. 
Taking the angular average over the kinetic equation \eqref{eq:sec4-W-Markov-eom-det} or \eqref{eq:sec4-W-Markov-eom} and using conditions \eqref{eq:sec4-zero-mode-constraints}, we obtain
\begin{align}
u\cdot \mathcal{D}[\mathcal{A}_r] W_0
+
\mathcal{D}_\mu[\mathcal{A}_r] \int_v n^\mu W_\perp(v)
=0 \,.
\label{eq:sec4-W0-continuity}
\end{align}
This is the local continuity equation for the hard sector color density. 
Therefore a collision kernel consistent with the hard-current Ward identity cannot directly damp the velocity average $W_0$. 
The collisions may relax the direction-dependent response $W_\perp$, but the averaged component $W_0$ should remain as part of the local charge dynamics.

We now take the short-memory limit only for $W_\perp$, to exhibit how the nonlocal hard response reduces to Markovian local transport. Collisions can make $W_\perp$ a fast angular response, while $W_0$ is tied to color-charge conservation and is kept explicitly.
This limit gives a collisional Markov regime and differs from the collisionless HTL limit. 
We assume that the collision kernel has a dissipative gap $\Gamma_{\rm gap}$ on the $P_\perp$ sector. 
The soft fields are taken to vary slowly compared with this relaxation scale
\begin{align}
\epsilon_{\rm sm}
\sim
\frac{\omega_{\rm cov}}{\Gamma_{\rm gap}},\ 
\frac{k_{\rm cov}}{\Gamma_{\rm gap}}
\ll 1 \,.
\label{eq:sec4-short-memory-hierarchy}
\end{align}
Here, $\omega_{\rm cov}$ and $k_{\rm cov}$ are their shorthand power-counting scale for covariant time and spatial variation, respectively. 
More explicitly, $\omega_{\rm cov}$ denotes the typical size of $u\cdot\mathcal D[\mathcal A_r]$ acting on the slowly varying fields, and $k_{\rm cov}$ denotes the typical size of $\Delta_u^{\mu\nu}\mathcal D_\nu[\mathcal A_r]$ acting on them. 
In this regime the direction-dependent response $W_\perp$ relaxes on the time scale $\Gamma_{\rm gap}^{-1}$ and can be solved locally as an expansion in $\epsilon_{\rm sm}$, whereas $W_0$ is kept explicitly.

To display the leading local term, we use the simplest isotropic truncation of the projected collision kernel
\begin{align}
C_{vv'}
=
\Gamma_{\rm iso}P_{\perp,vv'}
+
\cdots\,,
\qquad
\Gamma_{\rm iso}>0 \,.
\label{eq:sec4-isotropic-collision}
\end{align}
This ansatz means that all zero-mode-free angular components relax with the same rate $\Gamma_{\rm iso}$, while the velocity zero mode is left untouched. 
In this truncation, $\Gamma_{\rm iso}$ is the explicit representative of $\Gamma_{\rm gap}$, and the memory time of the fast angular response is of order $\Gamma_{\rm iso}^{-1}$. 
The omitted terms represent more general anisotropic or higher-angular-moment structures that preserve the same zero mode and contribute only subleading corrections in the short-memory expansion.

Projecting the kinetic equation onto $P_\perp$ and solving it to leading order in $\epsilon_{\rm sm}$, we find
\begin{align}
W_\perp(x,v)
=
\frac1{\Gamma_{\rm iso}}
\left[
v \cdot \mathcal{E}_r(x)
-
n^\nu \mathcal{D}_\nu[\mathcal{A}_r] W_0(x)
\right]
+
\mc{O}\!\left(\epsilon_{\rm sm}^2\right) \,.
\label{eq:sec4-Wperp-short-memory}
\end{align}
The first term is the local electric response of the fast angular modes. The second term is required because the slow charge mode $W_0$ remains present and is transported covariantly. Dropping this term would give an Ohmic response, but not a current satisfying the hard-current Ward identity.
Substituting \eqref{eq:sec4-Wperp-short-memory} into the current decomposition \eqref{eq:sec4-current-zero-mode-decomp} gives
\begin{align}
\mathcal J_{{\rm hard},{\rm det}}^\mu
=
m_D^2u^\mu W_0
+
\sigma_{\rm hard}
\left(
\mathcal{E}_r^\mu
-
\Delta_u^{\mu\nu}\mathcal D_\nu[\mathcal A_r]W_0
\right)
+
\mc{O}\!\left(m_D^2\epsilon_{\rm sm}^2\right)\,,
\quad
\sigma_{\rm hard}=\frac{m_D^2}{(d-1)\Gamma_{\rm iso}}\, .
\label{eq:sec4-local-conductivity-limit}
\end{align}
This is the local conductivity limit of the retarded hard current. The term proportional to $\mathcal E_r^\mu$ is the Ohmic part. The charge-storage term $m_D^2u^\mu W_0$ and the gradient correction involving $\mathcal D W_0$ are required by the Ward identity. With the zero-mode equation \eqref{eq:sec4-W0-continuity}, it is found that the complete short-memory current satisfies
\begin{align}
\mathcal{D}_\mu[\mathcal{A}_r] \mathcal{J}^\mu_{\rm hard,det}=0\,.
\label{eq:sec4-short-memory-Ward-complete}
\end{align}

\subsection{Noise, positivity, and thermal matching}
\label{subsec:sec4-noise-kms}

This subsection adds the fluctuation sector associated with the retained hard response. 
In the SK language, this comes from the part of the local action quadratic in $a$-type fields.
In the local $W$-sector action \eqref{eq:sec4-local-W-action}, this role is played by the kernel $N^{\rm loc}_{vv'}$. 
Thus $N^{\rm loc}_{vv'}$ is first a local noise kernel for the retained hard response $W_r(x,v)$. 
It is local in spacetime and acts on the same velocity labels as $W_r$. 
It should not be confused with the current-level symmetric kernel that appears only after $W$ is integrated out.

The zero-mode structure is fixed by the discussion in Sec.~\ref{subsec:sec4-retarded-projection}. The decomposition \eqref{eq:sec4-W0-Wperp} separates the velocity average $W_0$ from the direction-dependent response $W_\perp$. The current decomposition \eqref{eq:sec4-current-zero-mode-decomp} identifies $m_D^2W_0$ as the hard-sector color-density component, and \eqref{eq:sec4-W0-continuity} gives its continuity equation. Therefore the collisional and stochastic parts may act on $W_\perp$, but they should not directly relax or generate the velocity average $W_0$.

For the collision kernel, this means that the Ward identity requires
\begin{align}
\int_v C_{vv'}=0 \,.
\label{eq:sec4-C-left-zero-mode}
\end{align}
In the collisional Markov regime considered here, we take the projected form in which collisions act only on the direction-dependent sector. Writing $C$ as the integral operator associated with $C_{vv'}$
\begin{align}
(Cf)(x,v)
\coloneqq
\int_{v'}C_{vv'}f(x,v') \,,
\label{eq:sec4-C-operator-def}
\end{align}
this condition is
\begin{align}
\int_v C_{vv'}=0,
\qquad
\int_{v'}C_{vv'}=0,
\qquad
C=P_\perp C P_\perp .
\label{eq:sec4-C-zero-mode-both}
\end{align}
Here $P_\perp$ is the projector onto the direction-dependent zero-mode-free sector defined in \eqref{eq:sec4-P0-Pperp}. 
The first condition is required by the hard-current Ward identity. The full projected form says that the collision kernel does not act on $W_0$ and only relaxes the angular response $W_\perp$.

The local noise kernel satisfies the same zero-mode condition:
\begin{align}
\int_v N^{\rm loc}_{vv'}=0\,,
\qquad
\int_{v'}N^{\rm loc}_{vv'}=0\,,
\qquad
N^{\rm loc}=P_\perp N^{\rm loc}P_\perp\, .
\label{eq:sec4-local-noise-zero-mode}
\end{align}
Equivalently, the stochastic force in \eqref{eq:sec4-W-Markov-eom} has no velocity average. 
Thus the local noise may randomize the direction-dependent hard response, but it cannot directly change $W_0=\int_v W_r$.
The remaining conditions specify a stable collisional Markov sector for the retained variable $W$. 
These conditions are not consequences of gauge covariance alone.
They are physical inputs specifying that the collision term relaxes the direction-dependent response rather than amplifying it. 
From Eq.~\eqref{eq:sec4-K-operator}, the homogeneous equation for the fast angular response has the following schematic form
\begin{align}
\partial_t W_\perp + C W_\perp =0 \,.
\end{align}
For a time-independent collision kernel, this leads to, schematically,
\begin{align}
W_\perp(t)\sim \ee^{-Ct}W_\perp(0)\,.
\end{align}
Thus the dissipative part of the projected collision kernel is non-negative on the $P_\perp$ sector. 
In the color-neutral velocity-space approximation used here, we define
\begin{align}
(C_{\rm diss})_{vv'}
\coloneqq
\frac12\left(
C_{vv'}+C_{v'v}
\right)\,,
\label{eq:sec4-C-diss}
\end{align}
and impose
\begin{align}
C_{\rm diss}\ge0
\quad
\text{on }P_\perp\, .
\label{eq:sec4-C-positive}
\end{align}
For example, the isotropic choice $C_{vv'}=\Gamma_{\rm iso}P_{\perp,vv'}$ with $\Gamma_{\rm iso}>0$ satisfies this condition.

The local noise kernel describes a positive noise covariance on the same zero-mode-free sector. 
In the simple classical normalization used here, thermal matching to the stable Markovian collision sector is expressed as
\begin{align}
N^{\rm loc}
=\frac{2T}{m_D^2}
C_{\rm diss}
\quad
\text{on } P_\perp \,.
\label{eq:sec4-local-FDT-simple}
\end{align}
This is the local Markov form of the fluctuation-dissipation relation. 
It ties the noise strength to the same dissipative operator that governs the relaxation of $W_\perp$. 
Therefore, together with the stability condition $C_{\rm diss}\geq0$ in \eqref{eq:sec4-C-positive}, it implies the SK-positivity condition
\begin{align}
N^{\rm loc}\ge0
\quad
\text{on }P_\perp .
\label{eq:sec4-local-noise-positive}
\end{align}
Thus the local noise is not an independent nonlocal input. It is the thermal noise paired with the dissipative relaxation of the retained hard response, and the local fluctuation-dissipation relation is consistent with both stability of the collision sector and SK positivity of the local noise.\footnote{More generally, for the zero-mode-free Langevin equation $\partial_t W_\perp+C W_\perp=\xi_W$, let $\langle\cdots\rangle_{\rm eq}$ denote the expectation value in the local thermal equilibrium ensemble of the retained hard sector, and define the equal-time covariance by $S_W(v,v')\coloneqq\langle W_\perp(t,v)W_\perp(t,v')\rangle_{\rm eq}$. Equilibrium means that this covariance is time independent. Taking $dS_W/dt=0$ gives $0=-CS_W-S_WC^\dagger+N^{\rm loc}$, hence $N^{\rm loc}=CS_W+S_WC^\dagger$. For $S_W=T\,P_\perp/m_D^2$, this reduces to $N^{\rm loc}=\frac{2T}{m_D^2}
C_{\rm diss}$ on $P_\perp$.}
This is the velocity space analogue of the local color noise relation discussed in App.~\ref{app:dKMS-color-noise}.

So far, all kernels have referred to the local $W$ sector. 
We now pass to the noise seen by the soft gauge field after $W$ is integrated out. 
The point is that the local noise $N^{\rm loc}$ is not itself the soft current noise. 
It first propagates through the local $W$ equation and is then mapped to the hard current. 
Since the hard current is linear in $W$, we introduce the map
\begin{align}
(MW)^\mu(x)
\coloneqq
m_D^2\int_v v^\mu W(x,v)\,,
\label{eq:sec4-current-projection-map}
\end{align}
which is just rewrite of the local current \eqref{eq:sec4-hard-current-W} as a linear map.
After $W$ is integrated out, the local $W$-sector noise is propagated by the retarded and advanced Green functions of the $W$ equation and then mapped to the current. 
The resulting current-level noise kernel is
\begin{align}
N_J
=
M G_R^W N^{\rm loc} G_A^W M^\dagger\,,
\label{eq:sec4-noise-factorized}
\end{align}
where $G_R^W$ is the retained-response Green function defined in \eqref{eq:sec4-GR-def}, and $G_A^W$ is its advanced adjoint. 
Eq.~\eqref{eq:sec4-noise-factorized} is therefore not an additional nonlocal noise input but the current noise induced by the local $W$-sector noise after the retained hard response has been integrated out.

The soft influence functional is written directly in terms of current-level kernels. In a homogeneous background, the Fourier representation of $N_J$ gives the symmetric current kernel that appears in the quadratic soft action. 
We write the retarded current kernel by $G_R^{\mu\nu}$ and the symmetric current kernel by $G_{\rm sym}^{\mu\nu}$. 
Around a homogeneous color-neutral state, the quadratic part of the soft influence functional after the $W$ integration is
\begin{align}
I_{\rm hard,soft}^{(2)}
=
-2\int_k
\Tr\left[
A_{a\mu}(-k)G_R^{\mu\nu}(k;u)A_{r\nu}(k)
\right]
+
\iu\int_k
\Tr\left[
A_{a\mu}(-k)G_{\rm sym}^{\mu\nu}(k;u)A_{a\nu}(k)
\right]
+
\cdots \,,
\label{eq:sec4-quadratic-hard-effective-action}
\end{align}
where $\int_k\coloneqq\int\dd^dk/(2\pi)^d$, and the kernels are normalized so that $2\Tr(A_aJ)=A_a^AJ^A$ in components. In the collisional Markovian case, $G_{\rm sym}^{\mu\nu}$ is generated from the local kernel $N^{\rm loc}$ through propagation by $G_R^W$, $G_A^W$, and the current map $M$, as encoded in \eqref{eq:sec4-noise-factorized}.

As in the general SK discussion of Sec.~\ref{sec:SK-formalism}, thermal equilibrium requires the current-level kernels to obey the KMS relation
\begin{align}
G_{\rm sym}^{\mu\nu,AB}(k;u)
=
\coth\frac{\beta\omega_u}{2}\,
\frac{1}{2\iu}
\left[
G_R^{\mu\nu,AB}(k;u)-G_A^{\mu\nu,AB}(k;u)
\right]\,,
\qquad
\omega_u\coloneqq -u\cdot k\, .
\label{eq:sec4-KMS}
\end{align}
The advanced kernel is fixed by SK reality. In a homogeneous color-neutral background,
\begin{align}
G_A^{\mu\nu,AB}(k;u)
=
\left[
G_R^{\nu\mu,BA}(k;u)
\right]^*
\label{eq:sec4-GA-reality}
\end{align}
up to the standard momentum-conservation convention.

The collisionless HTL benchmark should be distinguished from this collisional Markov construction. When $C=0$, there is no local dissipative Markov operator and no associated local bulk noise, so $N^{\rm loc}=0$. This does not mean that the thermal symmetric HTL correlator vanishes. Rather, the source of the symmetric sector changes. In the collisional Markov case, the input fluctuation is the local bulk noise $N^{\rm loc}$ acting throughout the bulk evolution. In the collisionless case, there is no such bulk noise. The fluctuation is instead specified by the initial hard distribution and is then transported by the collisionless equation. Let $W_{\rm in}(x,v)$ denote the initial hard response data and $S_{W,{\rm in}}$ denote its covariance,
\begin{align}
S_{W,{\rm in}}(x,v;y,v')
\coloneqq
\left\langle
W_{\rm in}(x,v)W_{\rm in}(y,v')
\right\rangle_{\rm eq} .
\label{eq:sec4-initial-hard-covariance}
\end{align}
In the collisionless theory, this initial covariance plays the role of $N^{\rm loc}$ in the collisional Markov theory. 
Thus the collisionless counterpart of \eqref{eq:sec4-noise-factorized} becomes
\begin{align}
N_{J,{\rm collisionless}}
=
M G_R^{W,0} S_{W,{\rm in}} G_A^{W,0} M^\dagger\,,
\label{eq:sec4-collisionless-boundary-noise}
\end{align}
where $G_R^{W,0}$ is the collisionless retarded Green function of $v\cdot\mathcal D[\mathcal A_r]$, and $G_A^{W,0}$ is its advanced adjoint.
In a homogeneous thermal state, $S_{W,{\rm in}}$ is chosen so that the current-level kernels obey \eqref{eq:sec4-KMS}. 
Thus the collisionless HTL benchmark is deterministic in the bulk retarded sector, while its thermal Keldysh sector is supplied by initial hard-sector data rather than by local bulk noise.

\subsection{Collisionless HTL benchmark in physical variables}
\label{subsec:sec4-HTL-benchmark}

In this subsection, we take the collisionless retarded limit of the local hard-response construction and match it to the standard HTL induced current.
The symmetric sector in the collisionless theory was discussed in Sec.~\ref{subsec:sec4-noise-kms}; here we focus on the deterministic retarded current. 
The point is to show that the usual nonlocal HTL current is obtained by solving the local equation for the retained hard response $W_r(x,v)$ and then writing the result in ordinary YM variables.

The collisionless retarded benchmark is obtained by removing the bulk collision kernel
\begin{align}
C_{vv'}=0 \,.
\label{eq:sec4-collisionless-local-data}
\end{align}
In addition, if the local bulk noise is also absent 
\begin{align}
    N^{\rm loc}_{vv'}=0\,,    
\end{align}
the thermal symmetric sector is supplied instead by the initial hard distribution as described in Sec.~\ref{subsec:sec4-noise-kms}. 
The deterministic local equation for $W_r$ is then given by
\begin{align}
v\cdot\mathcal D[\mathcal A_r] W_r(x,v)
=
v\cdot\mathcal{E}_r(x) .
\label{eq:sec4-collisionless-W-eom-hframe}
\end{align}
Solving this equation with the retarded Green function, we obtain the current written in the $h_r$-dressed variables by
\begin{align}
\mathcal J_{\rm HTL}^\mu[\mathcal A_r](x)
=
m_D^2\int_v v^\mu
\left[\frac{1}{v\cdot\mathcal D[\mathcal A_r]}\right]_{\rm ret}
v\cdot\mathcal{E}_r(x)\,.
\label{eq:sec4-HTL-current-hframe}
\end{align}
This is the collisionless version of the retarded current derived in Sec.~\ref{subsec:sec4-retarded-projection}: the Green function $G_R^W$ is replaced with the retarded inverse of $v\cdot\mathcal D[\mathcal A_r]$.

The same current can be written in ordinary YM variables. 
We recall that the electric field is given by 
\begin{align}
E_r^\mu = F_r^{\mu\nu}u_\nu,
\qquad
u_\mu E_r^\mu=0 ,
\label{eq:sec4-physical-electric-field}
\end{align}
and that $D[A_r]$ is used for the ordinary covariant derivative. 
The physical HTL induced current is
\begin{align}
J_{\rm HTL}^\mu[A_r](x)
=
m_D^2\int_v v^\mu
\left[\frac{1}{v\cdot D[A_r]}\right]_{\rm ret}
v\cdot E_r(x).
\label{eq:sec4-HTL-current-induced-hard}
\end{align}
Equivalently, if we define
\begin{align}
W_{\rm HTL}(x,v)
\coloneqq
\left[\frac{1}{v\cdot D[A_r]}\right]_{\rm ret}
v\cdot E_r(x)\,,
\qquad
v\cdot D[A_r]W_{\rm HTL}=v\cdot E_r \,,
\label{eq:sec4-W-HTL-def}
\end{align}
then the HTL induced current is also written as 
\begin{align}
J_{\rm HTL}^\mu=m_D^2\int_v v^\mu W_{\rm HTL}\,.
\end{align}
The field $W_{\rm HTL}$ in \eqref{eq:sec4-W-HTL-def} is not a new retained variable but the retarded solution of the local variable $W_r$ after $W_r$ has been integrated out.

We now check that \eqref{eq:sec4-HTL-current-induced-hard} reproduces the standard retarded HTL polarization tensor. Linearizing around a homogeneous color-neutral background, the source $v\cdot E_r$ is first order in $A_r$.
The gauge-connection part of $v\cdot D[A_r]$ then gives only higher-order terms, because it acts on $W_{\rm HTL}=\mathcal O(A_r)$. Thus, for the quadratic response, the covariant inverse reduces to the retarded inverse of $v\cdot\partial$ as
\begin{align}
\left[\frac{1}{v\cdot D[A_r]}\right]_{\rm ret}
v\cdot E_r
=
\left[\frac{1}{v\cdot\partial}\right]_{\rm ret}
v\cdot E_r
+
\mathcal O(A_r^2)\,.
\end{align}
With the Fourier expansion of the gauge field
\begin{align}
A_{r\mu}^A(x)
=
\int_k \ee^{-\iu k\cdot x}A_{r\mu}^A(k),
\label{eq:sec4-Fourier-convention}
\end{align}
the retarded inverse gives
\begin{align}
\left[\frac{1}{v\cdot\partial}\right]_{\rm ret}
\longrightarrow
\frac{\iu}{v\cdot k+\iu0^+}.
\end{align}
At the same order, we find
\begin{align}
v\cdot E_r^A(k)
=
-\iu
\left[
(v\cdot k)u^\nu-(u\cdot k)v^\nu
\right]
A_{r\nu}^A(k),
\end{align}
where the sign follows from the Fourier convention above. Combining these two factors in \eqref{eq:sec4-HTL-current-induced-hard}, the induced current takes the form of
\begin{align}
J_{\rm HTL}^{\mu,A}(k)
=
\Pi_R^{\mu\nu}(k;u)A_{r\nu}^A(k),
\label{eq:sec4-HTL-polarization-def}
\end{align}
with
\begin{align}
\Pi_R^{\mu\nu}(k;u)
=
m_D^2\int_v v^\mu
\frac{(v\cdot k)u^\nu-(u\cdot k)v^\nu}{v\cdot k+\iu0^+} .
\label{eq:sec4-HTL-polarization}
\end{align}
The denominator is therefore the Fourier-space form of the retarded propagation along the hard trajectory.

This tensor is transverse.
The right transversality is immediate from the numerator,
\begin{align}
\Pi_R^{\mu\nu}(k;u)k_\nu=0 .
\end{align}
For the left transversality, multiplying \eqref{eq:sec4-HTL-polarization} by $k_\mu$ gives a factor $(v\cdot k)/(v\cdot k+\iu0^+)$. With the retarded prescription this factor is replaced by one in the distributional sense relevant for the HTL kernel, and the normalized velocity average $\int_v v^\mu=u^\mu$ gives
\begin{align}
k_\mu\Pi_R^{\mu\nu}(k;u)
=
m_D^2\int_v
\left[
(v\cdot k)u^\nu-(u\cdot k)v^\nu
\right]
=0 .
\end{align}
Equivalently, this tensor satisfies
\begin{align}
k_\mu\Pi_R^{\mu\nu}(k;u)=0\,,
\qquad
\Pi_R^{\mu\nu}(k;u)k_\nu=0 \,.
\label{eq:sec4-HTL-polarization-transverse}
\end{align}
In the local plasma rest frame, \eqref{eq:sec4-HTL-polarization} is the standard angular representation of the retarded HTL polarization tensor, written in the sign convention of \eqref{eq:sec4-HTL-current-induced-hard}.

The nonlinear Ward identity follows from the same velocity-average argument used in Sec.~\ref{subsec:sec4-velocity-markov}. Applying $D_\mu[A_r]$ to \eqref{eq:sec4-HTL-current-induced-hard} and using
\begin{align}
v\cdot D[A_r]W_{\rm HTL}=v\cdot E_r,
\end{align}
we find that
\begin{align}
D_\mu[A_r]J_{\rm HTL}^\mu
=
m_D^2\int_v v\cdot E_r
=
0 \,,
\label{eq:sec4-HTL-Ward}
\end{align}
because $u\cdot E_r=0$ and $\int_v n^\mu=0$. Therefore the collisionless retarded HTL current is gauge-covariant and satisfies the covariant Ward identity. The full thermal HTL influence functional also requires the symmetric sector discussed in Sec.~\ref{subsec:sec4-noise-kms}.

The retarded inverse in \eqref{eq:sec4-HTL-current-induced-hard} can be written explicitly as propagation along the hard trajectory. For a matrix-valued source $f$, this inverse acts as
\begin{align}
\left[\frac{1}{v\cdot D[A_r]}\right]_{\rm ret}f(x)
=
\int_0^\infty\dd s\,
U_r(x,x-vs)f(x-vs)U_r^{-1}(x,x-vs)\,,
\label{eq:sec4-retarded-inverse-wilson-line}
\end{align}
where $U_r(x,y)$ is the adjoint parallel transporter of the ordinary YM connection along the trajectory from $y$ to $x$. 
This formula only makes the retarded memory in \eqref{eq:sec4-HTL-current-induced-hard} explicit: the current at $x$ depends on the past source along the hard trajectory. 
In the local EFT, this memory is generated by integrating out $W_r$, not by inserting a bilocal current interaction from the start.

For completeness, we also record how this ordinary YM expression is related to the $h_r$-dressed variables used in the local system-environmental construction. 
If the retarded inverse is first written with the dressed connection $\mathcal A_r[h_r]$, the ordinary YM transporter is obtained by endpoint rotations,
\begin{align}
U_r(x,y)
=
h_r(x)\,
\mathcal U_{\mathcal A[h_r]}(x,y)\,
h_r^{-1}(y) \,.
\label{eq:sec4-h-frame-transporter-to-physical}
\end{align}
This relation should be read after $W_r$ has been integrated out while $h_r$ is still kept explicitly. 
If one wants a purely soft YM kernel, one may also integrate out the $h_r$ sector. 
Then $h_r$ is replaced by its retarded solution $h_{r,{\rm ret}}$ introduced in Eq.~\eqref{eq:sec3-h-ret-def}, which gives
\begin{align}
U_r^{\rm soft}(x,y)
=
h_{r,{\rm ret}}(x)\,
\mathcal U_{\mathcal A[h_{r,{\rm ret}}]}(x,y)\,
h_{r,{\rm ret}}^{-1}(y) .
\label{eq:sec4-soft-physical-transporter}
\end{align}
Thus integrating out $W_r$ produces the HTL current, while integrating out $h_r$ is an additional step needed only for a purely soft YM kernel.

The collisionless HTL current is therefore the retarded benchmark of the local system-environment construction. Its nonlocality is the retarded propagation of the retained hard response $W_r(x,v)$. The corresponding thermal symmetric sector is not generated by local bulk noise in the collisionless theory; it is supplied by the initial hard-sector data described in Sec.~\ref{subsec:sec4-noise-kms}.

\section{Summary and discussion}
\label{sec:conclusion}

We have constructed a local system-environment EFT for semiclassical non-Abelian gauge dynamics as the extension of the dissipative gravitational open EFT~\cite{Lau:2024mqm}.
The basic idea is to retain the environmental response that carries color charge, rather than integrating it out from the outset and writing a nonlocal influence functional directly. 
In the enlarged description, the dynamics is local and Markovian in the system and retained environmental variables. 
Nonlocal structures, such as memory kernels and Wilson-line factors, arise only after the retained response is solved using retarded Green functions and substituted back into the system-only description. 
In this sense, the construction follows the general logic of SK open-system EFTs and influence functionals \cite{Schwinger:1960qe,Keldysh:1964ud,Feynman:1963fq,Crossley:2015evo,Glorioso:2017fpd,Liu:2018kfw,Lau:2024mqm}, while its gauge-theoretic completion is closer in spirit to recent open EFT analyses of gauge systems \cite{Salcedo:2024nex,Kaplanek:2026kpp}.

A central ingredient is the environmental color frame and its $a$-type partner. 
These variables allow the environmental response to be written in a local color basis and make the Ward identity manifest. 
The color frame alone, however, is not a dissipative environment. 
The color-frame environment should also include current degrees of freedom that store and transport the color charge exchanged with the YM sector. 
We showed that such a current may contain susceptibility, conductivity, relaxation, and local noise, and that an isolated Ohmic term is not by itself compatible with color-charge conservation. 
After the retained environmental response is eliminated using retarded Green functions, the reduced system description contains deterministic memory currents and induced stochastic currents. 
Thus, gauge-covariant nonlocal kernels are not assumed as fundamental bilocal terms but they are generated from a local enlarged EFT. 
The same logic applies to matter channels. 
The local environmental damping written in the color frame becomes a gauge-covariant retarded self-energy in physical variables after the environmental response is eliminated.

As a concrete benchmark, we applied the framework to hard-loop response. 
By retaining a velocity-resolved adjoint response variable $W_r(x,v)$, the hard sector is described, before projection, by a local covariant transport equation. 
Integrating out $W_r$ at fixed color frame produces a retarded hard-loop memory kernel.
In the collisionless limit, this reproduces the standard HTL induced current, with the Wilson-line structure arising from the retarded inverse of $v\cdot D_r$. 
This connects the local Markov embedding developed here to the standard effective-action and kinetic-theory descriptions of hard thermal loops \cite{Braaten:1989mz,Frenkel:1989br,Kelly:1994dh,Blaizot:1993zk,Blaizot:2001nr}. 
The real-time fluctuation and noise structure of hard loops is naturally described in the SK formulation \cite{CaronHuot:2007nw}. 
In a Markov completion with collisions, the collision kernel and noise should preserve the velocity-space zero mode so that the hard current satisfies the Ward identity.
The short-memory limit then gives a local conductivity expansion with an explicit charge-storage mode, rather than an isolated nonconserved Ohmic current.

The distinctive advantage of the enlarged formulation appears when the environmental response is not parametrically short-lived. In this regime, integrating out the response generates nontrivial frequency dependence and memory that cannot be captured by a controlled finite-order derivative expansion of a local system-only EFT. Retaining the corresponding slow variables instead preserves the relevant non-Markovian response while keeping the theory local in the enlarged state space. 
In the case of non-Abelian gauge theory, the transferred color charge, the total color Ward identity, the protected velocity-space zero mode, and the relation between dissipation and noise remain explicit, rather than being hidden inside nonlocal Wilson-line memory kernels.

This recasts the evolution as a coupled local initial-value problem, amenable in principle to standard local time-integration methods and potentially useful as a starting point for real-time lattice implementations that keep the relevant gauge constraints explicit. Related auxiliary-field structures have been used in hard-loop simulations and gauge-covariant real-time lattice studies \cite{Rebhan:2008uj,Rebhan:2008ky,Ipp:2018hai,Hoshina:2020gdy}. We trade an enlarged state space for a controlled, local, and symmetry-preserving treatment of dynamically relevant memory.

A first natural extension is to develop open EFTs for QCD matter in which the environmental sector explicitly carries the exchanged color charge and energy-momentum, with the hard-loop example studied here providing a natural starting point.
More generally, the same architecture can be extended beyond the color-current sector to QCD media in which the environment also carries energy-momentum, fluid velocity, temperature, viscous response, and thermal fluctuations. 
This is precisely where an EFT description is useful: the system-environment architecture is universal at the level of symmetries and consistency conditions, while the low-energy response variables and matching data encode the microscopic realization. 
The QGP provides a prototypical example, but the appropriate retained variables and matching data depend on the phase and background under consideration. 
The color sector is naturally related to earlier descriptions of colored particles and non-Abelian fluid dynamics \cite{Wong:1970fu,Bistrovic:2002jx,Fernandez-Melgarejo:2016xiv}, while the energy-momentum sector should be completed in the spirit of modern dissipative hydrodynamic EFTs and gravitational open EFTs \cite{Crossley:2015evo,Glorioso:2017fpd,Liu:2018kfw,Lau:2024mqm}. 
Since the construction can be formulated in a background-diffeomorphism-covariant language, it may also be applied to expanding or accelerating plasma backgrounds \cite{Bjorken:1982qr,Gubser:2010ze,Gubser:2010ui,Akamatsu:2016llw}, background metric perturbations, or genuinely curved spacetimes, as in previous studies of high-temperature QCD coupled to gravity \cite{Brandt:1993wu,Brandt:1994mv}. 
This provides a natural starting point for gauge- and diffeomorphism-covariant open EFTs of QCD matter, in particular for nonstationary QGP dynamics and, more broadly, for possible applications to dense matter in compact-star environments \cite{Baym:2017whm,Alford:2007xm}.

Another useful direction is to clarify the relation between the present color-frame Markov embedding and the BRST-complete SK formulation of open non-Abelian gauge theories \cite{Kaplanek:2026kpp}. 
The latter provides a quantum parent framework for BRST-compatible influence functionals, whereas the present construction gives a local bottom-up parametrization of the semiclassical response layer. 
Establishing the dictionary between these descriptions, including the roles of $h_\pm$, $\pi_a$, $W_r(x,v)$, retarded elimination, and Keldysh kernels, would provide a useful consistency check and may help connect the present dissipative EFT to quantum real-time applications.

The broader lesson is that nonlocal open-system response in gauge theory can be organized as the projection of a local system--environment theory in which the relevant memory channels are retained explicitly. This viewpoint separates symmetry completion from microscopic matching data and provides a systematic framework for constructing dissipative and stochastic non-Abelian gauge dynamics compatible with symmetry, causality, and positivity.

\section*{Acknowledgments}
\noindent
The work of YA was supported by JST Grant No. JPMJPF2221.

\appendix
\section{Dynamical KMS completion of dissipative color response}
\label{app:dKMS-color-noise}

This appendix gives a local dynamical KMS completion of the dissipative color response used in Sec.~\ref{sec:color-fluid-completion}.  
Dynamical KMS is the SK implementation of the KMS periodicity of a thermal state: in a local effective theory one combines microscopic time reversal with a thermal translation along the local thermal vector, and requires the SK action to be invariant under the resulting transformation \cite{Crossley:2015evo,Glorioso:2016gsa,Glorioso:2017fpd,Liu:2018kfw}.  The construction below is local in the retained system-environment EFT, before the retained color response is eliminated into a projected nonlocal system-only kernel.

The purpose is limited.  We do not derive the existence of the dissipative color channel, nor the numerical value of its conductivity, from symmetry alone.  Rather, once this local transverse color-response channel is retained, we show that dynamical KMS fixes the paired $a$-$a$ noise kernel, and SK positivity fixes the sign of the dissipative conductivity.

We first recall the $h$-dressed YM source.  
On the two SK branches, we define
\begin{align}
\mathcal A_{\pm\mu}
=
h_\pm^{-1}A_{\pm\mu}h_\pm
-
\frac{\iu}{\gc}h_\pm^{-1}\partial_\mu h_\pm \,.
\label{eq:app-dressed-source}
\end{align}
Near the diagonal SK saddle, we use the corresponding $r/a$ variables
\begin{align}
\mathcal A_{r\mu}
=
\frac12
\left(
\mathcal A_{+\mu}
+
\mathcal A_{-\mu}
\right)\,,
\qquad
\mathcal A_{a\mu}
=
\mathcal A_{+\mu}
-
\mathcal A_{-\mu}\,.
\end{align}
In this appendix, we impose the dynamical KMS transformation on the dressed source $\mathcal A_{\pm\mu}[A_\pm,h_\pm]$.  Equivalently, the transformation below should be understood as the induced dynamical KMS transformation of the gauge-covariant $h$-dressed source.
In the $h_r$ frame, the local color chemical potential and color electric field are
\begin{align}
\mu_h
&\coloneqq
u^\mu\mathcal A_{r\mu}\,,
&
\mathcal E_r^\mu
&\coloneqq
\mathcal F_r^{\mu\nu}u_\nu\,,
&
u_\mu\mathcal E_r^\mu
&=0 \,.
\label{eq:app-mu-E-def}
\end{align}
The dressed field strength satisfies $\mc{F}_{r\mu\nu} = h_r^{-1} F_{r\mu\nu} h_r$.

Before constructing the dynamical KMS variation, it is useful to separate the two parts of the retained color current with respect to the local environmental velocity as 
\begin{align}
\mathcal J_h^\mu
=
\mathcal Q_h u^\mu
+
\mathcal V_h^\mu\,,
\qquad
u_\mu\mathcal V_h^\mu=0\, .
\label{eq:app-current-decomposition}
\end{align}
The first term is the charge-storage part.  
For example, locally one may have
\begin{align}
\mathcal Q_h
=
\chi_h\mu_h+\cdots\, ,
\end{align}
where $\chi_h$ is a static susceptibility.  The dissipative conductivity and its local noise kernel instead belong to the transverse transport part:
\begin{align}
\mathcal V_h^\mu
=
\sigma_{h,{\rm diss}}^{\mu\nu}\mathfrak F_{h\nu}
+\cdots ,
\qquad
u_\mu\sigma_{h,{\rm diss}}^{\mu\nu}
=
\sigma_{h,{\rm diss}}^{\mu\nu}u_\nu
=
0 \,.
\label{eq:app-transverse-current-response}
\end{align}
This is why the dynamical KMS variation relevant for the noise-dissipation relation is the transverse part of the thermal variation of the dressed source.

We now construct that local thermal variation.  The first thermal datum is the local thermal vector defined by
\begin{align}
\beta^\mu
\coloneqq
\frac{u^\mu}{T}\,.
\end{align}
The second one is a thermal gauge twist, which is the gauge-theoretic version of the internal rotation in a charged thermal density matrix,
\begin{align}
\rho_{\beta,\mu}
\sim
\exp[-\beta(H-\mu Q)]\, .
\end{align}
The thermal circle is then accompanied not only by imaginary-time translation generated by $H$, but also by an internal rotation generated by $Q$.  The dimensionless parameter of this rotation is $\beta\mu=\mu/T$. 
In gauge theory, this internal rotation is represented as a gauge transformation around the thermal circle.  
We write the corresponding matrix-valued twist in the $h_r$ frame by
\begin{align}
\nu_h
\coloneqq
\frac{\mu_h}{T}\,.
\label{eq:app-thermal-twist}
\end{align}
More invariantly, $\nu_h$ is the homogeneous adjoint thermal twist; if one introduces an explicit thermal gauge parameter, $\nu_h$ is the covariant combination of that parameter with $\beta^\mu\mathcal A_{r\mu}$.  In the local $h_r$ frame of the retained color-diffusion sector, this twist is identified with $\mu_h/T$.

For an ordinary tensor, the real part of the dynamical KMS shift is generated
by the Lie derivative along the thermal vector $\beta^\mu$.  
The dressed source $\mathcal A_{r\mu}$ is already invariant under the original physical gauge transformation.  
Nevertheless, it is a connection in the local $h_r$-color frame, and the Lie derivative of a connection naturally splits into a horizontal, adjoint-covariant part and a vertical frame-rotation part.
With the conventions of Sec.~\ref{sec:relative-gauge-frame}, this kinematic
identity reads
\begin{align}
\pounds_\beta\mathcal A_{r\mu}
-
\mathcal D_\mu[\mathcal A_r]
\left(
\beta^\rho\mathcal A_{r\rho}
\right)
=
\beta^\rho\mathcal F_{r\rho\mu}\,.
\label{eq:app-covariant-Lie}
\end{align}
It is noted that the subtraction in the left-hand side should not be viewed as restoring the original physical gauge invariance, which is already built into $\mathcal A_{r\mu}$ rather than as removing the connection, or vertical, piece generated by dragging the local color frame along $\beta^\mu$.
The result is the $h_r$-frame adjoint-covariant one-form $\beta^\rho\mathcal F_{r\rho\mu}$.

We write the dynamical KMS shift added to the $a$-type dressed source as $\delta_{\beta,\nu_h} \mc{A}_{a\mu}$.
Including the thermal color twist, we define
\begin{align}
\delta_{\beta,\nu_h}\mathcal A_{a\mu}
&\coloneqq
\pounds_\beta\mathcal A_{r\mu}
-
\mathcal D_\mu[\mathcal A_r]
\left(
\beta^\rho\mathcal A_{r\rho}
\right)
+
\mathcal D_\mu[\mathcal A_r]\nu_h
= \beta^\rho \mc{F}_{r\rho\mu} + \mc{D}_\mu[\mc{A}_r]\nu_h\, .
\label{eq:app-thermal-variation-long}
\end{align}
where \eqref{eq:app-covariant-Lie} is used.\footnote{In the Abelian, constant-temperature local-rest-frame limit, this thermal variation reduces to the standard diffusion dynamical KMS variation reviewed in Sec.~5 of Ref.~\cite{Liu:2018kfw}.  In the present notation, the spatial component becomes $\delta_{\beta,\nu}\mathcal A_{ai}=\beta_0\partial_t\mathcal A_{ri}$, with $\beta_0=1/T$.}
Thus the local thermal shift is the sum of the covariant thermal translation of the $h_r$-frame connection and the thermal color twist.
We note that in the classical $\hbar$-counting of the SK-EFT, this shift is counted as $a$-type.
On the other hand, the transformation law of the $r$-type variables is defined in the standard way:
\begin{align}
\delta_{\beta,\nu_h}\mathcal A_{r\mu}
=0\,.
\label{eq:app-thermal-variation-r}
\end{align}

The sign of the twist term follows the convention
\begin{align}
\rho_{\beta,\mu}
\sim
e^{-\beta(H-\mu Q)}\,,
\qquad
\mu_h=u^\mu\mathcal A_{r\mu}\,.
\end{align}
With this convention, the transverse thermal variation reduces in a local rest frame to
\begin{align}
-\frac1T
\left[
\mathcal E_i-T\mathcal D_i
\left(
\frac{\mu_h}{T}
\right)
\right]\,.
\end{align}
Changing the sign of the twist term would correspond to the opposite convention for the chemical potential or for the electrochemical force.

We now project this thermal variation onto the transverse diffusion sector:
\begin{align}
\Phi_{h\mu}
\coloneqq
\Delta_{u\mu}{}^\nu
\delta_{\beta,\nu_h}\mathcal A_{a\nu}\,.
\label{eq:app-Phi-def}
\end{align}
This is not an additional dynamical assumption; it selects the same sector in which the color conductivity $\sigma_{h,{\rm diss}}$ and the noise kernel $N_h$ act.  
Eqs.~\eqref{eq:app-thermal-twist} and \eqref{eq:app-thermal-variation-long} makes \eqref{eq:app-Phi-def} rewritten as 
\begin{align}
\Phi_{h\mu}
=
\Delta_{u\mu}{}^\nu
\left[
\beta^\rho\mathcal F_{r\rho\nu}
+
\mathcal D_\nu[\mathcal A_r]
\left(
\frac{\mu_h}{T}
\right)
\right]\,.
\label{eq:app-Phi-expanded}
\end{align}
Using the relations,
\begin{align}
\beta^\rho
=
\frac{u^\rho}{T}\,,
\qquad
\mathcal E_{r\mu}
=
\mathcal F_{r\mu\nu}u^\nu\,,
\qquad
u^\rho\mathcal F_{r\rho\mu}
=
-\mathcal E_{r\mu}\,,
\end{align}
we obtain
\begin{align}
\Phi_{h\mu}
=
-\frac1T\mathcal E_{r\mu}
+
\Delta_{u\mu}{}^\nu
\mathcal D_\nu[\mathcal A_r]
\left(
\frac{\mu_h}{T}
\right)
=
-\frac1T
\left[
\mathcal E_{r\mu}
-
T\Delta_{u\mu}{}^\nu
\mathcal D_\nu[\mathcal A_r]
\left(
\frac{\mu_h}{T}
\right)
\right]\,.
\label{eq:app-Phi-force-step}
\end{align}
This motivates the transverse electrochemical force given by
\begin{align}
\mathfrak F_h^\mu
\coloneqq
\mathcal E_r^\mu
-
T\Delta_u^{\mu\nu}
\mathcal D_\nu[\mathcal A_r]
\left(
\frac{\mu_h}{T}
\right)\,,
\qquad
u_\mu\mathfrak F_h^\mu=0\, ,
\label{eq:app-color-force}
\end{align}
so that
\begin{align}
\Phi_{h\mu}
=
-\frac1T\mathfrak F_{h\mu}\,.
\label{eq:app-Phi-force}
\end{align}
Thus the electrochemical force is not inserted by hand into the dynamical KMS shift.  It appears as a kinematic identity: the transverse part of the gauge-covariant thermal variation of the $h_r$-dressed source is $-\mathfrak F_h/T$.

We use the standard time-reversal parities for the transverse current sector.  These parities are part of the definition of the local time-reversal operation, not EOM.  
In the local rest frame, we define
\begin{align}
\mathcal A_{a\mu}^{\perp}
\coloneqq
\Delta_{u\mu}{}^\nu\mathcal A_{a\nu}\,,
\end{align}
which is the spatial source coupled to a spatial current and is therefore odd under time reversal.  
The anti-unitary time-reversal operation $\Theta$ acts as 
\begin{align}
\Theta\mathcal A_{a\mu}^{\perp}
=
-\mathcal A_{a\mu}^{\perp}\,,
\qquad
\Theta\Phi_{h\mu}
=
\Phi_{h\mu}\,,
\qquad
\Theta\mathfrak F_{h\mu}
=
\mathfrak F_{h\mu}\,.
\label{eq:app-parity}
\end{align}

The classical dynamical KMS transformation then has the form
\begin{align}
\widetilde{\mathcal A}_{a\mu}^{\perp}
=
\Theta\mathcal A_{a\mu}^{\perp}
+
\iu\,\Theta\Phi_{h\mu}
+\cdots\, ,
\label{eq:app-dKMS-before-parity}
\end{align}
where the omitted terms are higher order in the derivative and $a$-field expansion.  Using \eqref{eq:app-parity} and \eqref{eq:app-Phi-force}, we obtain
\begin{align}
\widetilde{\mathcal A}_{a\mu}^{\perp}
=
-\mathcal A_{a\mu}^{\perp}
+
\iu\Phi_{h\mu}
=
-\mathcal A_{a\mu}^{\perp}
-
\frac{\iu}{T}\mathfrak F_{h\mu}\,.
\label{eq:app-local-dKMS}
\end{align}
The first term is the time-reversal parity of the transverse current source; the second is the dynamical KMS shift by the thermal variation of the $r$-type dressed source.

We now impose \eqref{eq:app-local-dKMS} on the local quadratic action for the dissipative transverse response and its noise.  Since all kernels below act on the transverse sector, we suppress the explicit superscript $\perp$ on $\mathcal A_a$ from now on.  The relevant local terms are
\begin{align}
I_{2,{\rm loc}}
=
2\int\dd^dx\,
\Tr\left[
-
\mathcal A_{a\mu}
\sigma_{h,{\rm diss}}^{\mu\nu}
\mathfrak F_{h\nu}
+
\frac{\iu}{2}
\mathcal A_{a\mu}
N_h^{\mu\nu}
\mathcal A_{a\nu}
\right]
+\cdots \,.
\label{eq:app-local-action}
\end{align}
Here $\sigma_{h,{\rm diss}}^{\mu\nu}$ is the symmetric dissipative part of the local color conductivity, and $N_h^{\mu\nu}$ is the symmetric local noise kernel.  Both kernels are transverse:
\begin{align}
u_\mu\sigma_{h,{\rm diss}}^{\mu\nu}
=
\sigma_{h,{\rm diss}}^{\mu\nu}u_\nu
=
0\,,
\qquad
u_\mu N_h^{\mu\nu}
=
N_h^{\mu\nu}u_\nu
=
0 \,.
\label{eq:app-transverse-kernels}
\end{align}
Possible antisymmetric transverse responses are nondissipative transport data and are not involved in the positivity argument below.

For the coefficient comparison, let us define
\begin{align}
(X,KY)
\coloneqq
2\int\dd^dx\,
\Tr\left(
X_\mu K^{\mu\nu}Y_\nu
\right)\,,
\label{eq:app-bilinear}
\end{align}
and write \eqref{eq:app-local-action} as 
\begin{align}
I_{2,{\rm loc}}
=
-
(\mathcal A_a,\sigma_{h,{\rm diss}}\mathfrak F_h)
+
\frac{\iu}{2}
(\mathcal A_a,N_h\mathcal A_a)
+\cdots \,.
\label{eq:app-local-action-bilinear}
\end{align}
Under \eqref{eq:app-local-dKMS}, the dissipative term transforms as
\begin{align}
-
(\mathcal A_a,\sigma_{h,{\rm diss}}\mathfrak F_h)
\longmapsto
-
\left(
-\mathcal A_a-\frac{\iu}{T}\mathfrak F_h,
\sigma_{h,{\rm diss}}\mathfrak F_h
\right)
=
(\mathcal A_a,\sigma_{h,{\rm diss}}\mathfrak F_h)
+
\frac{\iu}{T}
(\mathfrak F_h,\sigma_{h,{\rm diss}}\mathfrak F_h)\,.
\label{eq:app-dKMS-diss-transform}
\end{align}
The noise term transforms as
\begin{align}
\frac{\iu}{2}
(\mathcal A_a,N_h\mathcal A_a)
&\longmapsto
\frac{\iu}{2}
\left(
-\mathcal A_a-\frac{\iu}{T}\mathfrak F_h,
N_h
\left[
-\mathcal A_a-\frac{\iu}{T}\mathfrak F_h
\right]
\right)
\nonumber\\
&=
\frac{\iu}{2}
(\mathcal A_a,N_h\mathcal A_a)
-
\frac1T
(\mathcal A_a,N_h\mathfrak F_h)
-
\frac{\iu}{2T^2}
(\mathfrak F_h,N_h\mathfrak F_h)\,.
\label{eq:app-dKMS-noise-transform}
\end{align}
The mixed term in the transformed action is therefore
\begin{align}
\left(
\mathcal A_a,
\left[
\sigma_{h,{\rm diss}}
-
\frac1T N_h
\right]
\mathfrak F_h
\right)\,.
\label{eq:app-transformed-mixed}
\end{align}
Dynamical KMS invariance requires this mixed term to reproduce the original mixed term
\begin{align}
-
(\mathcal A_a,\sigma_{h,{\rm diss}}\mathfrak F_h).
\end{align}
Hence, the noise kernel and the color conductivity should satisfy
\begin{align}
\sigma_{h,{\rm diss}}
-
\frac1T N_h
=
-\sigma_{h,{\rm diss}}\,,
\end{align}
and therefore they are relate by
\begin{align}
N_h
=
2T\,\sigma_{h,{\rm diss}}\,.
\label{eq:app-FDT-bilinear}
\end{align}
Restoring spacetime and adjoint indices gives
\begin{align}
N_h^{\mu\nu}
=
2T\,\sigma_{h,{\rm diss}}^{\mu\nu}\,.
\label{eq:app-FDT}
\end{align}
The same relation also cancels the pure force term generated by the dynamical KMS shift:
\begin{align}
\frac{\iu}{T}
(\mathfrak F_h,\sigma_{h,{\rm diss}}\mathfrak F_h)
-
\frac{\iu}{2T^2}
(\mathfrak F_h,N_h\mathfrak F_h)
=
0
\qquad
\text{when } N_h=2T\sigma_{h,{\rm diss}}\,.
\end{align}

For an isotropic color-neutral environment, the color structure is proportional to the adjoint identity and the tensor structure reduces to
\begin{align}
\sigma_{h,{\rm diss}}^{\mu\nu}
=
\sigma_h\Delta_u^{\mu\nu},
\qquad
N_h^{\mu\nu}
=
N_h\Delta_u^{\mu\nu}.
\end{align}
Then \eqref{eq:app-FDT} becomes
\begin{align}
N_h
=
2T\sigma_h,
\qquad
N_h^{\mu\nu}
=
2T\,\sigma_h\,\Delta_u^{\mu\nu}\,.
\label{eq:app-isotropic-FDT}
\end{align}

Finally, SK positivity fixes the sign of the dissipative conductivity.  Since the path-integral weight is $\exp(\iu I_{\rm eff})$, the local noise term contributes a damping factor of the form
\begin{align}
\exp\left[
-\frac12
(\mathcal A_a,N_h\mathcal A_a)
\right]\,.
\end{align}
Thus $N_h$ must define a non-negative quadratic form:
\begin{align}
(B,N_hB)
=
2\int\dd^dx\,
\Tr\left(
B_\mu N_h^{\mu\nu}B_\nu
\right)
\ge0
\label{eq:app-noise-positivity}
\end{align}
for every transverse adjoint-valued test field $B_\mu$.  Combining this with \eqref{eq:app-FDT-bilinear}, and assuming $T>0$, gives
\begin{align}
(B,\sigma_{h,{\rm diss}}B)
=
2\int\dd^dx\,
\Tr\left(
B_\mu\sigma_{h,{\rm diss}}^{\mu\nu}B_\nu
\right)
\ge0 \,.
\label{eq:app-sigma-positivity}
\end{align}
In the isotropic color-neutral case this becomes simply
\begin{align}
\sigma_h\ge0 \,.
\end{align}

The conclusion is therefore as follows.  The force $\mathfrak F_h^\mu$ is the electrochemical force of the chosen local dissipative color channel.  Its appearance in the dynamical KMS shift is not an independent assumption: it follows from the transverse projection of the gauge-covariant thermal variation of the $h_r$-dressed source, including the thermal gauge twist $\nu_h=\mu_h/T$.  The transverse projection is used because $\sigma_h$ and its paired noise kernel belong to the spatial transport sector $\mathcal V_h^\mu$, not to the charge-storage sector $\mathcal Q_hu^\mu$.  Once this local dissipative channel is retained, dynamical KMS fixes the paired noise kernel by
\begin{align}
N_h^{\mu\nu}
=
2T\,\sigma_{h,{\rm diss}}^{\mu\nu}\,,
\end{align}
in the classical local limit, and SK positivity requires $\sigma_{h,{\rm diss}}$ to be positive semi-definite.  The susceptibility $\chi_h$ is different in character: it is a static matching coefficient subject to thermodynamic stability, whereas $\sigma_h$ is a dissipative transport coefficient constrained by the dynamical KMS and positivity argument above.

\newcommand{\arxivfont}{\rmfamily}
\bibliographystyle{yautphysm}
\bibliography{ref}

\end{document}